\documentclass{article}

\usepackage[english]{babel}
\usepackage[letterpaper,top=1cm,bottom=2cm,left=2cm,right=2cm,marginparwidth=1.75cm]{geometry}

\usepackage{amsmath}
\usepackage{graphicx}
\usepackage{graphicx}%
\usepackage{multirow}%
\usepackage{amsmath,amssymb,amsfonts}%
\usepackage{amsthm}%
\usepackage{mathrsfs}%
\usepackage[title]{appendix}%
\usepackage{xcolor}%
\usepackage{textcomp}%
\usepackage{manyfoot}%
\usepackage{booktabs}%
\usepackage{algorithm}%
\usepackage{algorithmicx}%
\usepackage{algpseudocode}%
\usepackage{listings}%

\usepackage{comment}
\usepackage{balance}
\usepackage{array}
\usepackage{xcolor}
\usepackage{enumitem}
\usepackage{subcaption}
\usepackage{tabularx}
\usepackage{algorithm}
\usepackage{algorithmicx}
\usepackage{titlesec}
\usepackage{algpseudocode}
\usepackage{caption}
\usepackage[most]{tcolorbox}
\usepackage{graphicx}
\usepackage{booktabs}
\usepackage{threeparttable}
\usepackage{siunitx}
\usepackage{booktabs}
\usepackage{siunitx}
\usepackage{makecell}
\usepackage{multirow}
\usepackage{array}
\newcolumntype{L}[1]{>{\raggedright\let\newline\\\arraybackslash\hspace{0pt}}m{#1}}
\newcolumntype{C}[1]{>{\centering\let\newline\\\arraybackslash\hspace{0pt}}m{#1}}
\newcolumntype{R}[1]{>{\raggedleft\let\newline\\\arraybackslash\hspace{0pt}}m{#1}}

\usepackage{hyperref}
\hypersetup{
  colorlinks=true,
  citecolor=blue,
  linkcolor=purple,
  urlcolor=blue
}

\newcommand{\keywords}[1]{%
  \vspace{0.5em}
  \noindent\textbf{Keywords: }#1
}

\newcommand{\requiredPeerCount}{\textit{requiredPeerCount}}
\newcommand{\maxPeerCount}{\textit{maxPeerCount}}

\title{End-to-End and Phase-Level Performance Optimization for\\
Hyperledger Fabric}

\author{
Pavan Sollu$^{2}$,
Aniruddha Mukherjee$^{1}$,
Divya Pulivarthi$^{1}$,
S.R. Eshwar$^{1}$,\\
Gugan Thoppe$^{1}$,
Kshitij Pratihast$^{2}$,
Tittu Varghese$^{2}$,
Hrishikesh Nashikkar$^{2}$,\\
Yogesh Simmhan$^{1*}$\\[0.8em]
$^{1}$Indian Institute of Science (IISc), Bangalore, India.\\
$^{2}$National Payments Corporation of India (NPCI), Hyderabad, India.\\[2em]
$^{*}$Corresponding author(s). E-mail(s): \href{mailto:simmhan@iisc.ac.in}{simmhan@iisc.ac.in};
}

\date{}

\begin{document}

\maketitle

\begin{abstract}
Hyperledger Fabric (HLF) is a modular, permissioned blockchain widely adopted in enterprise settings where privacy, controllable trust and predictable performance are paramount. Enhancing its throughput and latency remains challenging, as optimization decisions made in one phase of the transaction lifecycle can adversely affect other phases.
In this work, we present a systematic, phase-level and end-to-end study of HLF optimizations along three fronts, combining production-grade testbed experiments with calibrated SimPy simulations. First, we introduce two novel optimization techniques that target commit-phase bottlenecks -- \textit{block-level pipelining} and \textit{strategic waiting}. 
In pipelining, we overlap validation and private-data acquisition of successive blocks with state-consistency checks and ledger updates
improving commit throughput by up to $1.9\times$. 
Strategic waiting coordinates commit progress by temporarily pausing fast leaders and boosting laggers to sustain endorsement parallelism, 
yielding up to a $1.2\times$ higher throughput. Second, we conduct detailed micro-benchmarking of three key configuration levers: private-data dissemination, block-size selection, and endorsement peer selection. Our results reveal that: 
(i) Relaxed quorums for private-data dissemination significantly reduce latency in both endorsement and commit phases; 
(ii) Under light workloads, smaller blocks yield lower end-to-end latency, whereas, under heavy workloads, larger blocks are necessary to improve throughput and reduce latency; and 
(iii) Relaxed leader selection dramatically reduces dropped transactions and boosts endorsement throughput, with a modest increase in MVCC invalidations. 
Finally, we analyze the interplay among private-data dissemination, VSCC parallelization, and pipelined commits. Interestingly, the throughput gains over a serial commit path are maximized at a moderate level of parallelization (rather than extreme core provisioning). 
Together, our findings provide actionable, phase-aware guidance and protocol-level refinements for optimizing HLF deployments at scale.
\end{abstract}


\section{Introduction}
\label{sec:Introduction}

Blockchain is an immutable transaction ledger maintained across a distributed network of \textit{peers}. \textit{Public blockchains}~\cite{blockchain-public-private-permissioned} such as Bitcoin, Ethereum, FileCoin and Helium~\cite{Bitcoin, Ethereum, FileCoin, Helium} are open for participation,  and rely on consensus mechanisms such as Proof of Work (PoW) or Proof of Stake (PoS)~\cite{pow-pos-king2012ppcoin}, which can be resource intensive.
The included transactions may also in turn call \textit{smart contracts}~\cite{blockchain-smartcontracts}, which store and execute business logic, to enforce agreements when their conditions are met. 
\textit{Private and permissioned blockchains}~\cite{blockchain-public-private-permissioned} adapt these ideas for enterprise settings that require stronger access control, data confidentiality, and predictable performance, 
e.g., fintech \cite{IBM_WhitePaper, wu2024blockchainfinancesurvey} and healthcare~\cite{sensors-healthcare, agbeyangi2024blockchain}. Here, access to the blockchain is limited to one or more organizations, and consensus typically relies on more efficient protocols such as Raft~\cite{raft-paper}, enabling higher throughput and lower latency relative to public blockchains.
\textit{Hyperledger Fabric (HLF)}~\cite{first_hlf_paper}, an open-source project from the Linux Foundation, has emerged as the \textit{de facto} platform for enterprise-grade permissioned blockchains.

Given HLF's growing adoption, undertsanding and optimizing its end-to-end performance is critical. 
We ground our study in the context of the \textit{National Payments Corporation of India (NPCI)}, which operates UPI, one of the world's largest payment networks processing over 500 million transactions daily. While UPI relies on a centralized gateway for monetary exchange, emerging financial instruments such as the eRupee are being transacted and tracked using HLF. 
Although current volumes are modest ($\approx$1M transactions/day), they are projected to grow rapidly, motivating careful performance engineering to avoid future scalability bottlenecks. 

\begin{figure*}[!t]
    \centering
    \includegraphics[width=0.94\linewidth]{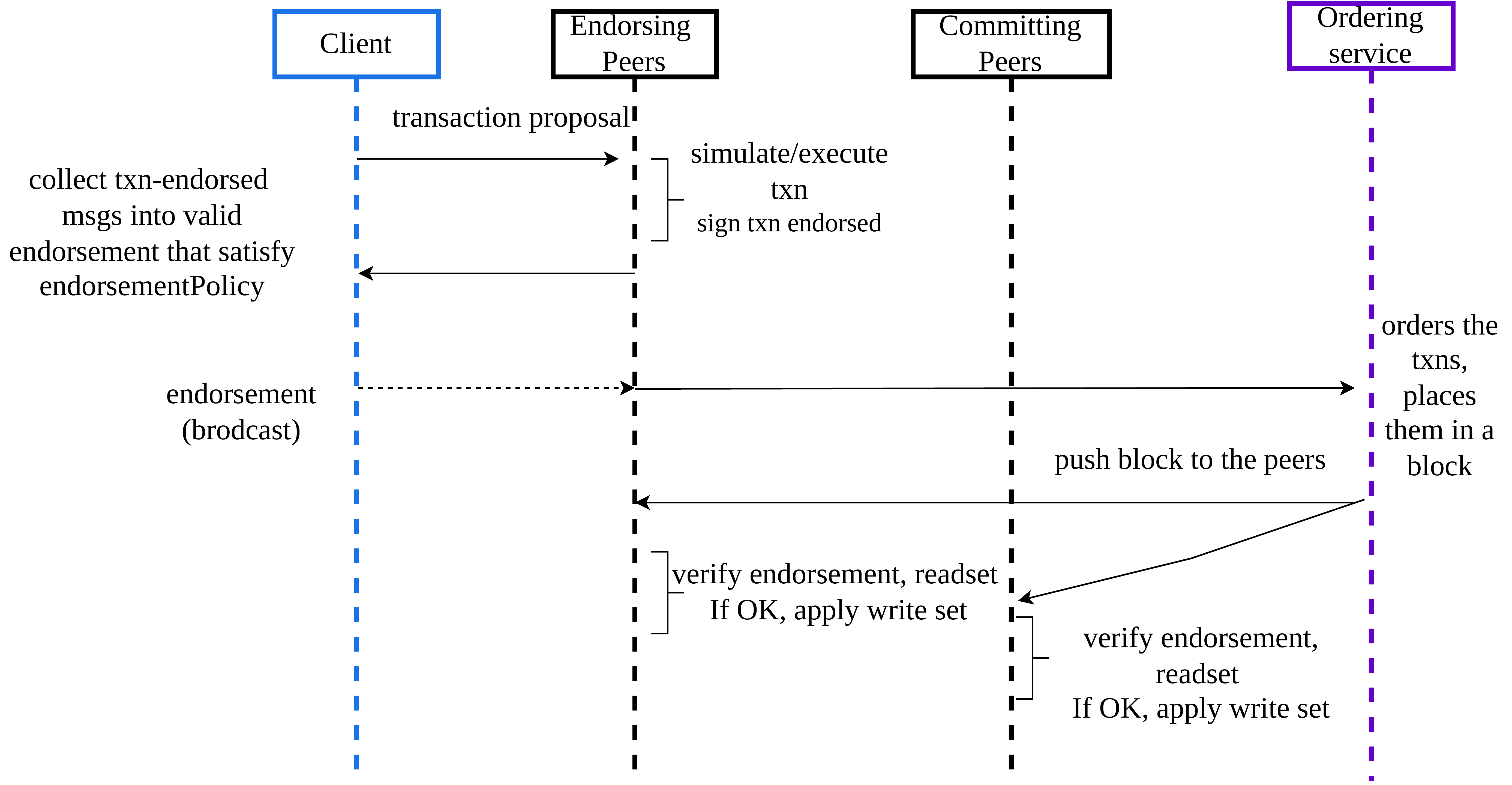}
    \caption{Transaction flow in HLF}
    \label{fig:hlf_transaction_flow}
\end{figure*}

In HLF, every transaction progresses through a well-defined lifecycle consisting of three phases: \textit{endorsement}, \textit{ordering}, and \textit{commit}. When a transaction is submitted to HLF for execution by a client, designated endorser peers simulate execution of the proposed transaction, do preliminary checks, and share any private data with other peers in the network that is necessary for downstream commits. After this endorsement, an ordering service sequences multiple endorsed transactions and packages them into blocks, ensuring a consistent global order across the network. These blocks are broadcast to all peers, who validate each transaction against the endorsement and data integrity rules, and append these blocks to their local ledgers and complete the commit phase. 
Figure~\ref{fig:hlf_transaction_flow} and Section~\ref{sec:HLF_Background} provide a detailed overview of this process.

Existing literature on HLF optimization propose parallel execution of independent transactions and smart contracts~\cite{baheti2022dipetrans, liu2021parallel}, benchmarking and performance analysis of HLF~\cite{perf-benchmarking-of-hlf-springer,6-page-saudi-shuaib2022performance,melo2024performance}, and optimizations for HLF~\cite{thakkar2018performance} such as early detection of conflicts~\cite{moving-mvcc-to-orderer}, alternative consensus protocols~\cite{10087049} and different databases~\cite{goleveldb-benchmark}.
However, most existing studies optimize or analyze individual phases in isolation, without systematically characterizing how decisions made in one phase interact with and influence others. 
This lack of phase-aware, end-to-end evaluation is a challenge. Locally beneficial choices can counterintuitively degrade overall system performance, e.g., larger blocks reduce per-block commit overhead under heavy load yet increase latency under light load.
Moreover, independently effective strategies may not be complementary, yielding diminished or even negative returns when combined, e.g., asynchronous private-data sharing to speed up endorsements and commit-stage tuning to reduce latency performed together do not offer net gains.

In this work, we address this gap through a systematic, phase-level and end-to-end analysis of HLF performance.
Our main contributions are: 
\begin{enumerate}[leftmargin=*]

    \item We use production-grade HLF runs and large-scale simulations for detailed \textit{phase-level micro-benchmarking} of configuration levers such as private-data dissemination, block-size tuning under varying loads, and leader selection. 
    Our experiments reveal the following insights. 
    \begin{enumerate}[label=(\roman*),wide = 0pt]

        \item \emph{Private-data dissemination}: We systematically benchmark three private-data dissemination strategies on HLF~v2.4.7 and show that endorsement latency is dominated by the acknowledgment quorum (182--227\,ms for single-ack vs.\ 275--360\,ms for all-ack), while commit latency is driven by data availability. Single-peer sharing~$(1,1)$ incurs expensive on-demand fetches ($>2000$\,ms), causing severe leader--straggler disparity, whereas broadcast strategies reduce fetch times to $\approx500$\,ms and equalize commit performance. Based on these results, we propose a relaxed quorum~$(n{-}1,1^*)$ that achieves single-ack endorsement speed with broadcast-level commit efficiency.

        \item \emph{Block-size selection}: We show that block-size choice must be load-aware. Under low load (100~TPS), throughput is insensitive to block size, but end-to-end latency grows nearly $4\times$ as block size increases from 500 to 2000 transactions, favoring smaller blocks. Under high load (800~TPS), larger blocks improve throughput and reduce latency initially (26\% drop from 500 to 1000 tx/block), but gains rapidly diminish, with latency and throughput saturating beyond 1000--1500 transactions.

        \item \label{enum:leader} \emph{Leader selection}: We evaluate four endorsement-selection policies and show that relaxing selection from a single highest block-height peer to a broader eligibility increases endorsement throughput by $\approx1.29\times$--$1.50\times$ and reduces block-creation time by 22--33\%, but also increases MVCC invalidations by $\approx2.1\times$--$3.5\times$. This highlights a fundamental trade-off between endorsement drops and commit-time conflicts.

    \end{enumerate}

\item Beyond characterization, we introduce two complementary optimizations: \textit{block-level pipelining} and \textit{strategic waiting}, that explicitly exploit phase dependencies to improve commit throughput and sustain endorsement parallelism.
        \begin{enumerate}[label=(\roman*),wide = 0pt]
            \item \textit{Block-level pipelining} is a novel two-phase commit strategy that overlaps state-consistency checks and ledger updates of one block with validation checks and private-data acquisition of the next, breaking HLF's one-block-at-a-time commit bottleneck and improving throughput by up to $1.9\times$.

            \item We identify that heterogeneous commit speeds destabilize Soft Max-Ht leader selection by collapsing it into a single-leader regime (1.\ref{enum:leader} above), and propose \emph{strategic waiting}, a coordination mechanism that preserves endorsement parallelism and improves throughput by up to $1.2\times$.
        \end{enumerate}

\item Lastly, we examine the impact of private-data broadcast, VSCC parallelization, and pipelining on the commit phase. Individually, each of these optimizations accelerates commits, with pipelining alone improving throughput by 1.3-1.5× over a serial path (see Table 10). Surprisingly, our results show that throughput gains over a serial commit path is maximized at a moderate level of VSCC parallelization (not at the extremes). Specifically, commit TPS rises steadily with core scaling from 881 (1 core) to 2584 (64 cores, 4-1 setup; 1.94×) but then saturates at 2667 (96 cores; 1.98×), showing that peak speedups occur at moderate-to-high but not extreme levels of parallelization.
\end{enumerate}

The rest of this paper is organized as follows. Section \ref{sec:HLF_Background} provides background information and outlines HLF's architecture. Section \ref{sec:RelatedWorks} presents a review of related work. Section \ref{sec:Methods} describes the proposed methodology, including our approaches to relaxed quorum for private data flush, block size selection, two-phase pipelining for block commits, criteria for selecting endorsement peers and strategic waiting. Section \ref{sec:Exp_And_Results} details the experiments conducted and the corresponding results. Finally, Section \ref{sec:Conclusion} concludes the paper and discusses directions for future research.



\section{Background on Hyperledger Fabric}
\label{sec:HLF_Background}

\subsection{Design of HLF}
\paragraph*{Entities} There are three key entities in the HLF architecture.

A \textbf{client} is an external application that interacts with the blockchain through the HLF SDK, generating transaction proposals that invoke or query chaincode functions as part of smart contracts.
These proposals are sent to a subset of the endorsing peers in the network, and the endorsed transactions are forwarded to the Orderer through a designated peer for inclusion in the blockchain.

\textbf{Peers} are responsible for both \textit{endorsing} transactions and \textit{committing} blocks, thereby ensuring consensus and integrity. 
Each peer maintains a local \textit{ledger} consisting of the \textit{blockchain}, a complete history of all transactions, and the \textit{world state}, a snapshot of the current values of all assets.

\textbf{Orderers} are dedicated peers which establish a total order over endorsed transactions, batch and order them into blocks according to block-cut parameters, such as block size or block timeout, and disseminate finalized blocks to peers for validation and commit.
The committer peers will commit the block to their blockchain after successful validation.

Importantly, in HLF, endorsing and committing roles are performed by the same peer processes, depending on the transaction phase.
Often, these peers span across multiple organizations, but we focus here on the case where the peers are part of the same organization.

\paragraph*{Concepts} There are several concepts related to peers that are relevant to our design. 
The \textit{block height} of a peer is the number of blocks it has committed and reflects how up-to-date its local ledger is relative to other peers.
Peers with the highest block height within an organization are referred to as \textit{leaders}, while peers with lower block heights are termed \textit{stragglers}.
While all peers are committers, the leader peers are typically used as endorsing peers as they maintain the latest confirmed state of the system. 
Stragglers cannot safely endorse transactions due to stale state and may accumulate large commit backlogs, degrading overall system throughput.
Peers in an organization discover other peers through a gossip protocol based on gRPC, and periodically exchange metadata such as block height, endpoint information, and liveliness status. This helps peers determine the most updated or healthy peers in the network for endorsement or data retrieval.

\paragraph*{Transaction Flow} 
HLF follows an \textit{Execute--Order--Validate (EOV)} transaction model (Figure~\ref{fig:hlf_transaction_flow}), enabling speculative execution prior to ordering and increasing concurrency compared to traditional order--execute designs.
This guarantees consistency across all participants but restricts performance and flexibility, since execution must be deterministic.
The execute-order-validate model used in HLF~\cite{Androulaki_2018} improves on this by allowing transactions to be executed speculatively before they are ordered. During speculative execution (done during endorsements), if a transaction accesses stale or invalid data, then it is discarded during the validation phase using simple deterministic checks. This design enhances parallel processing of transactions, increasing throughput.

The transaction flow begins when a client submits a transaction proposal to invoke a chaincode to a group of endorsers. 
During endorsement, peers simulate transaction execution and produce a \textit{read-write set} capturing the state reads and proposed updates~\cite{hlf-doc-read-write_set}. 
The endorsers sign the results and return them to the client, which assembles the transaction with the endorsement signatures and submits it to the ordering service. In parallel, the endorser shares any private data within that transaction with other authorized peers, to be used later for validation.

The orderers are responsible for establishing consensus on the sequence of transactions. 
Endorsed transactions are ordered into blocks by the ordering service and broadcast to all peers for validation and commitment~\cite{hlf-ordering-service}. 
The orderer nodes follow a Raft consensus algorithm~\cite{raft-paper} to order the transactions in a deterministic, but not necessarily the same order they are received in, and place them in a block.

During commit, committer peers perform \textit{VSCC validation} to verify endorsement policy compliance, ensure read-write consistency through \textit{MVCC checks}, retrieve any required private data, and update the local ledger state.
Private data may either be available at the committer (if shared during endorsement) or will need to be fetched from other peers that may have it. If all validation conditions are satisfied, the committer applies the transaction's write-set to its local blockchain and updates its StateDB. If not, the transaction is marked invalid but still recorded in the block for auditability.

\subsection{Optimization Challenges}
Optimizing HLF requires carefully balancing throughput and end-to-end latency across tightly coupled transaction phases.
Configuration choices that accelerate one phase, such as relaxed endorsement acknowledgment or smaller blocks, can shift costs to downstream phases, degrading overall performance.
For instance, endorsement of a transaction, which entails speculative execution and private data sharing, can be accelerated by increasing the number of leaders and relaxing acknowledgment requirements. However, a looser leader-selection criterion increases the risk of commit invalidations due to stale data, while a lower acknowledgment threshold heightens the chance of missing private data, thereby stalling the commit pipeline as peers fetch the missing information.

Similarly, the time a transaction spends in the ordering phase, where endorsed transactions are batched into blocks, can be reduced by using smaller block sizes or shorter timeouts. However, producing more blocks amplifies the block-processing overheads during commit.

Lastly, while VSCC validation can be parallelized, MVCC checks and ledger updates are inherently sequential, creating commit-phase bottlenecks under high load. 
Separately, private data retrieval can be delayed if there are multiple leaders
or if the data were not widely disseminated in order to reduce endorsement time.

These cross-phase dependencies make isolated, phase-specific optimization insufficient, motivating coordinated, end-to-end performance analysis.


\section{Related Work}\label{sec:RelatedWorks}

Prior work on Hyperledger Fabric (HLF) performance broadly falls into three categories: (i) Performance optimization, (ii) Benchmarking and empirical analysis, and (iii) Analytical and simulation-based modeling.

\subsection{Performance Optimization Strategies}

Much of the optimization literature targets isolated bottlenecks within the Execute--Order--Validate (EOV) transaction flow of HLF, including endorsement verification, VSCC parallelism, and state-database access. 
Thakkar et al.~\cite{thakkar2018performance} identify several key bottlenecks in early HLF versions, 
and propose optimizations such as Membership Service Provider (MSP) caching and parallel VSCC validation, which have been incorporated into mainstream HLF releases.
Others~\cite{Javaid-optimal-validation-phase} propose parallelizing phases within the commit cycle of a single block for improved performance.

However, these approaches leave key commit components -- most notably MVCC checks and ledger updates -- largely sequential, limiting achievable throughput under high load.
This results in under-utilized CPU resources (with vertical scaling) and increased redundancy (with horizontal scaling). 
Inter-block pipelining and sparse-peer execution have been explored to overcome these limits, but often rely on dependency tracking, buffering of speculative state, or workload partitioning that complicate deployment and degrade goodput under contention~\cite{ibm-pipeline-sparse-peers-thakkar}.

Another key challenge in HLF's performance is the high rate of transaction failures due to MVCC conflicts. One line of work tames such conflicts by reordering the transactions, doing dependency analysis with early aborts, and using orderer-side pre-validations \cite{sharma_fabric_pp,ruan-fabric-sharp, moving-mvcc-to-orderer}. 
There are also several works improving performance in blockchains beyond HLF~\cite{baheti2022dipetrans, tian2024slchain, reshi2024blockchain, xu2020flexible, vairagade2025blockchain, zhou2024blockchain, puppala2024optimizing, datta2025energy, puppala2025enhancing, vairagade2022secure}.

In contrast to these approaches, our work introduces block-level pipelining that preserves HLF's commit correctness constraints while avoiding dependency-graph construction or speculative state exposure.

\subsection{Performance Benchmarking \& Modeling}

A substantial body of work benchmarks HLF performance under varying block sizes, arrival rates, state databases, and chaincode configurations.
For instance,~\cite{thakkar2018performance, block-size-valuation-hlf, hang2021optimal-block-size, sharma_fabric_pp} show that small blocks lower latency at low load, while larger blocks amortize per-block overhead and raise throughput under heavy load. Similarly, there have been works that benchmark the performance effect of state databases~\cite{thakkar2018performance, hang2021optimal-block-size, goleveldb-benchmark}, chaincode languages~\cite{chaincode-perf-analysis, thakkar2018performance}, node count and peer heterogeneity~\cite{chaincode-perf-analysis, thakkar2018performance}, and endorsement policy design~\cite{alhabib2026evaluating}.

These studies establish well-known trends, such as small blocks favoring low-latency regimes and large blocks improving throughput under heavy load, but typically focus on aggregate, end-to-end metrics.
Fine-grained, phase-level analyses that capture endorsement, ordering, validation, and commit interactions remain limited, particularly for private-data dissemination and leader-selection policies.

Additionally, analytical models based on queuing theory~\cite{theoretical-hlf-modelling-queuing}, Stochastic Petri Nets and variants (for example, SRNs)~\cite{melo2024comprehensive, sukhwani-srn-perf-modelling, ke2023performance-spn}, and hierarchical models~\cite{jiang2020performance-hierarchial} that predict throughput, latency, and queue behavior have also been studied. 
Such theoretical models provide useful abstractions but necessarily simplify system interactions, often missing emergent behaviors and making them ill-suited for studying cross-phase effects that arise when multiple optimizations interact in real-world deployments.

\subsection{Research Gap}
Overall, existing work lacks a systematic, phase-aware, end-to-end evaluation of HLF that jointly considers endorsement behavior, commit-path structure, configuration choices, and peer heterogeneity.

Existing optimization approaches often exhibit performance limitations under realistic workloads. For instance, while both intra-block and inter-block pipelining have been investigated~\cite{Javaid-optimal-validation-phase, ibm-pipeline-sparse-peers-thakkar}, the former provides only a limited performance boost, whereas the latter suffers a sharp decline in goodput (valid transactions per second) under high transaction inter-dependencies. An effective approach that leverages inter-block pipelining without relying on explicit dependency-graph construction is notably absent from the literature. Similarly, while there are many works~\cite{Javaid-optimal-validation-phase, sharma_fabric_pp, ruan-fabric-sharp} that optimize isolated stages with HLF, they often overlook coordinated strategies that  simultaneously improve endorsements and validation, thereby enhancing end-to-end throughput.

On the benchmarking front, prior studies have performed detailed analyses on selected configuration parameters~\cite{geyer2019performance, Javaid-optimal-validation-phase}, yet a fine-grained, phase-level characterization of private-data dissemination mechanisms and leader-selection policies on end-to-end throughput remains unexplored. 
In particular, the interaction between private-data dissemination, endorsement peer selection, and commit-phase behavior has received little attention, despite its significant impact on throughput and latency.
Finally, the majority of existing studies evaluate older HLF versions, leaving a gap in the systematic benchmarking of the latest architecture. 
Our work addresses this gap by combining detailed phase-level benchmarking with coordinated optimization mechanisms that operate across endorsement and commit stages on recent HLF v2.x releases.


\section{Methods}\label{sec:Methods}
In Section~\ref{sec:Exp_And_Results}, we empirically evaluate three key HLF configuration levers: \textit{private data dissemination} (\S~\ref{methods:pvt-data}), \textit{block size selection} (\S~\ref{methods:block-size}), and \textit{endorsement peer selection} (\S~\ref{methods:leader-selecn}), as well as two proposed strategies: \textit{Pipelining} (\S~\ref{methods:pipelining}) and \textit{Strategic Waiting} (\S~\ref{methods:strat-wait}). 
In this section, we describe the design choices and experimental setup underlying these studies. For each lever or mechanism, we present its motivation, the performance question being addressed, and the expected system-level effects.

\subsection{Optimal Private Data Sharing Strategies}\label{methods:pvt-data}

\begin{table*}[t]
    \centering
    \small
    \caption{Hypothesized impact of Private Data Sharing Parameters on endorsement time, commit time, and transaction success rate.}
    \label{tab:private-data-sharing-configuration-values_assumptions}
    \begin{tabular}{c||c|c|c|c}
    \hline
        \bf Method & \bf $\maxPeerCount$  & \bf $\requiredPeerCount$ & \bf Endorse time & \bf Commit time \\ \midrule
        \bf HLF default & 1 & 1 & low & high  \\
        \bf HLF default & $n$ - 1 & 1 &  high & low  \\
        \bf HLF default & $n$ - 1 & $n$ - 1 & high & low  \\
        \bf Proposed & $n$ - 1 & $1\ast$ & \textbf{low} & \textbf{low}\\
    \hline
    \end{tabular}
\end{table*}

\textit{Motivation.~}
Private-data dissemination in HLF creates a fundamental tension between the endorsement and commit phases.
During endorsement, an endorsing peer shares the transaction's private data with other peers in the organization to enable downstream validation and commit.
Disseminating to many peers improves data availability at commit time, but the endorser must wait for acknowledgments before proceeding, which slows endorsement.
Conversely, sending to fewer peers keeps endorsement fast, but shifts the cost to the commit phase, where peers that lack the data must fetch it on-demand over the network.
Crucially, because these costs arise in different phases of the pipeline, the net end-to-end effect of a given strategy is not obvious: a policy that looks optimal when measured at one phase may degrade overall throughput when both phases are considered together.

\textit{Objective.~} Our objective is to systematically characterize this tradeoff by measuring endorsement latency, commit latency (broken down by sub-stage), and end-to-end throughput under each dissemination strategy.
 
\textit{Approach.~} 
HLF exposes two configuration parameters that govern private-data dissemination.
The first, \texttt{maxPeerCount}, specifies the maximum number of peers to which the endorser sends the private data. The second, \texttt{requiredPeerCount}, specifies the minimum number of acknowledgments the endorser must receive before it considers the endorsement successful. Together, these parameters define a dissemination strategy, which we denote as a tuple \texttt{(maxPeerCount, requiredPeerCount)}.

Using these knobs, we study three existing strategies that span the design space.
Under \textbf{(1,\,1)}, the endorser sends data to exactly one peer and waits for that peer's acknowledgment.
Under \textbf{(\textit{n}$-$1,\,\textit{n}$-$1)}, the endorser broadcasts to all peers and waits for all acknowledgments, maximizing availability but exposing endorsement to the slowest responder.
Under $(n{-}1,\,1)$, the endorser broadcasts private data to all peers but designates a specific peer whose acknowledgment determines success; however, the endorser still waits for all responses to arrive before verifying that this acknowledgment is present.
Despite requiring only one specific acknowledgment for success, this strategy remains bottlenecked by the slowest responder, since the endorser waits for all replies before checking whether the designated peer's acknowledgment is present.

These baseline strategies therefore expose an inherent coupling between endorsement progress and the slowest responding peer.
Motivated by the limitations of these baselines, we also propose a \textbf{relaxed-quorum} strategy, denoted \textbf{(\textit{n}$-$1,\,1$^{*}$)}. Here, the endorser broadcasts to all peers but proceeds as soon as the \emph{first} valid acknowledgment arrives from \emph{any} peer, decoupling progress from any single responder while retaining broad dissemination.
Table~\ref{tab:private-data-sharing-configuration-values_assumptions} summarizes the expected qualitative impact of each strategy.
 
\textit{Expected Impact.} We expect the $(1,1)$ strategy to yield the fastest endorsements but the slowest commits, since non-endorsing peers will be forced into expensive network fetches.
The full-broadcast strategies should invert this picture: slower endorsements but faster commits due to local data availability.
Among the broadcast variants, we expect $(n{-}1, 1^{*})$ will achieve endorsement latency comparable to the lightweight $(1,1)$ policy (since it needs only one acknowledgment and is not bottlenecked by a specific peer) while retaining the commit-side benefits of broad dissemination.
If this expectation holds, the relaxed quorum mitigates the endorsement--commit tradeoff rather than merely shifting cost between phases.


\subsection{Block Size Selection under Varying Loads}
\label{methods:block-size}
 
\textit{Motivation. }The block size, the number of endorsed transactions the orderer accumulates before cutting a block, introduces a fundamental tradeoff between transaction latency and throughput.
A small block is cut quickly, so transactions spend less time waiting in the orderer queue, but the system must then process many blocks, and each block incurs a fixed commit overhead (VSCC setup, ledger writes, state-database flushes).
A large block amortizes this per-block overhead across more transactions, but forces early-arriving transactions to wait until the block is full, increasing their end-to-end latency.
What makes this tradeoff non-trivial is that the dominant contributor to end-to-end latency shifts with load.
Under light load, the orderer queue is rarely full, so block-creation time (waiting for transactions to arrive) dominates end-to-end latency and commit overhead is negligible.
Under heavy load, transactions arrive faster than blocks can be committed, so commit time dominates and blocks queue up waiting for processing.
The same block size can therefore be optimal in one regime and harmful in another.

\textit{Objective. }Our objective is to quantify how block size interacts with arrival rate, measuring block-creation time, per-stage commit latency (across the five sequential commit stages shown in Figure~\ref{fig:hlf_validation_phases}), end-to-end latency, and throughput.
 
\textit{Approach. }
We define the following terms used throughout our block-size analysis.
\emph{Block-creation time} is the wall-clock duration from when the orderer begins accumulating transactions for a block until the block is cut and dispatched.
\textit{Block-commit time} is the time taken by a peer to process a received block through all five commit stages: VSCC validation, private-data fetch, MVCC checks, block-store update, and state-database commit.
\emph{End-to-end latency} is the time from when a transaction is first submitted by a client to when it is committed to the ledger.
When the commit rate cannot keep pace with the block creation rate, blocks accumulate in the commit queue, and the resulting \textit{queuing delay} becomes the dominant contributor to end-to-end latency.

We sweep block sizes across 500, 1000, 1500, and 2000 transactions per block under two load regimes: a low-load scenario (100~TPS per peer) and a high-load scenario (800~TPS per peer). 
For each configuration, we profile block-creation time and the per-stage breakdown of block-commit time on our production-grade HLF testbed, allowing us to identify which commit sub-stages dominate under different load and block-size regimes.

\textit{Expected Impact. }We expect two distinct regimes.
Under low load, we expect block-creation time to dominate end-to-end latency, growing roughly linearly with block size, while commit time remains small—making smaller blocks preferable in this regime.
Under heavy load, we expect the picture to invert: block-creation time collapses (since transactions arrive fast), but commit time rises and becomes the bottleneck.
In this regime, larger blocks are expected to amortize fixed per-block commit overhead, reduce the number of commit cycles, and improve overall throughput.

These effects are workload- and deployment-dependent, motivating adaptive block-size selection rather than a single static choice.


\subsection{Criteria for Selecting Endorsement Peers}\label{methods:leader-selecn}

\begin{table*}[t]
    \centering
    \footnotesize
    \caption{Endorsement Peer Selection Strategies: Hypothesized Trade-off Summary.
    $\Uparrow\Uparrow$\,/\,$\Downarrow\Downarrow$ denote extreme high/low;
    $\Uparrow$\,/\,$\Downarrow$ denote moderate high/low;
    $\Rightarrow$ denotes moderate.}
    \label{tab:endorsing-peer-selection}

    \begin{tabular}{c||c|c|c|c}
    \hline
        \bf Strategy
        & \bf \makecell{Stale State\\Endorsement}
        & \bf \makecell{Tx Drop\\Rate}
        & \bf \makecell{Conflict\\Rate}
        & \bf \makecell{Block Create\\Time} \\
    \midrule

        \bf Max-Ht (Strict)
        & $\Downarrow\Downarrow$
        & $\Uparrow\Uparrow$
        & $\Downarrow\Downarrow$
        & $\Uparrow\Uparrow$ \\

        \bf Soft Max-Ht (Controlled)
        & $\Downarrow$
        & $\Downarrow$
        & $\Rightarrow$
        & $\Rightarrow$ \\

        \bf Ranked List (Relaxed)
        & $\Uparrow$
        & $\Downarrow\Downarrow$
        & $\Uparrow$
        & $\Downarrow$ \\

        \bf All Peers (Unrestricted)
        & $\Uparrow\Uparrow$
        & $\Downarrow\Downarrow$
        & $\Uparrow\Uparrow$
        & $\Downarrow\Downarrow$ \\

    \hline
    \end{tabular}
\end{table*}

\textit{Motivation. }When a client submits a transaction, it must be routed to one or more peers for endorsement.
The choice of endorsement peers directly affects two competing metrics: endorsement throughput and commit-time transaction validity.
A strict policy that restricts endorsement to only the most up-to-date peers (those with the highest block height) ensures that endorsed transactions are based on fresh state, minimizing MVCC invalidations at commit time.
However, funneling all endorsements to a small set of peers risks overloading them, causing transaction drops when their buffers saturate.
Conversely, a permissive policy that allows any peer to endorse maximizes concurrency and reduces endorsement drops, but peers with stale ledger state may endorse transactions against outdated data, leading to MVCC conflicts that are detected only during the commit phase.
These late-stage invalidations are operationally more expensive than early endorsement drops, as they consume ordering, replication, and validation resources and permanently record invalid transactions in the ledger.
The net effect on end-to-end success -- the fraction of submitted transactions that are both endorsed and committed as valid -- depends on the interplay between these two failure modes and varies with the level of transaction inter-dependence.

\textit{Objective. }Our objective is to evaluate how each selection policy affects three metrics: the number of successfully endorsed transactions, the number of endorsement-stage drops, and the number of MVCC invalidations at commit time, across varying levels of transaction dependency.
 
\textit{Approach. }We define four endorsement peer selection strategies: Max-Ht, Soft Max-Ht, Ranked List and All, ordered from strictest to most permissive.
Under \textit{Max-Ht}, only the peer(s) with the absolute highest block height can endorse.
Under \textit{Soft Max-Ht}, any peer whose block height lies within a configurable threshold~$\tau$ of the maximum observed height is eligible.
Under \textit{Ranked List}, the default HLF mechanism, peers are sorted in decreasing order of block height and transactions are forwarded in round-robin fashion starting from the highest; if a peer's buffer is full, the transaction falls through to the next peer in the ranking.
Under \textit{All}, every peer is eligible regardless of block height.
We additionally parameterize transaction dependencies: each incoming transaction is made dependent on a prior in-flight transaction with probability~$p$, so that if the prior transaction has not yet committed, the dependent transaction will fail MVCC validation.
This allows us to sweep the contention spectrum from fully independent ($p = 0$) to fully dependent ($p = 1$) workloads.
Table~\ref{tab:endorsing-peer-selection} summarizes the expected qualitative tradeoffs associated with each endorsement-selection strategy.

We conduct a SimPy-based discrete-event simulation in which five peers, each backed by a client generating transactions at a fixed rate, process endorsements and commits with service times sampled from empirical latency distributions collected on our HLF testbed. Each peer maintains a bounded gateway buffer and a capped endorsement concurrency; arrivals that exceed both limits are dropped. We compare all four strategies across dependency probabilities $p \in \{0, 0.2, 0.4, 0.6, 0.8, 1.0\}$, recording endorsed counts, drop counts, MVCC invalidation counts, and average block-creation time for each configuration.
 
\textit{Expected Impact. }We expect Max-Ht to produce the fewest MVCC invalidations but the highest drop rate, since a single overloaded leader must absorb the entire endorsement load.
As we relax toward Soft Max-Ht and Ranked List, we expect drops to fall sharply as more peers absorb endorsement work, but MVCC invalidations should rise because some of those peers endorse against stale state.
All Peers is expected to eliminate endorsement drops but at the cost of substantially higher MVCC invalidations.


\subsection{Two-phase Pipelining for Block Commits}\label{methods:pipelining}

\begin{figure*}[t]
    \centering
    \includegraphics[width=\linewidth]{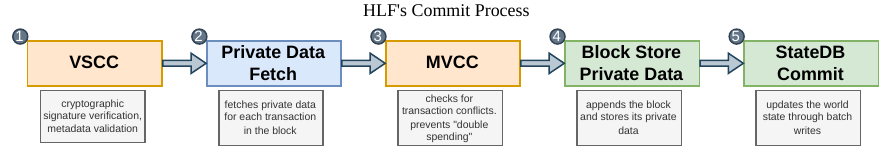}
    \caption{Stages of the commit phase in HLF validation.}
    \label{fig:hlf_validation_phases}
\end{figure*}

\begin{figure*}[t]
    \centering
    \includegraphics[width=1\linewidth]{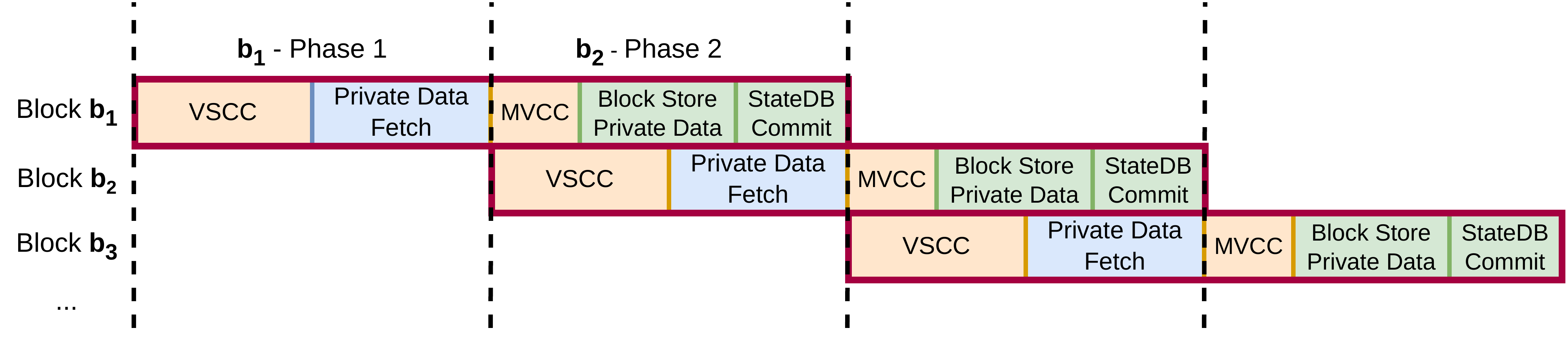}
    \caption{Pipelining the validation phase (ideal case)}
    \label{fig:commit_pipeline}
\end{figure*}

\begin{figure*}[t]
\centering
\begin{subfigure}[b]{0.48\textwidth}
\centering
\includegraphics[width=\linewidth]{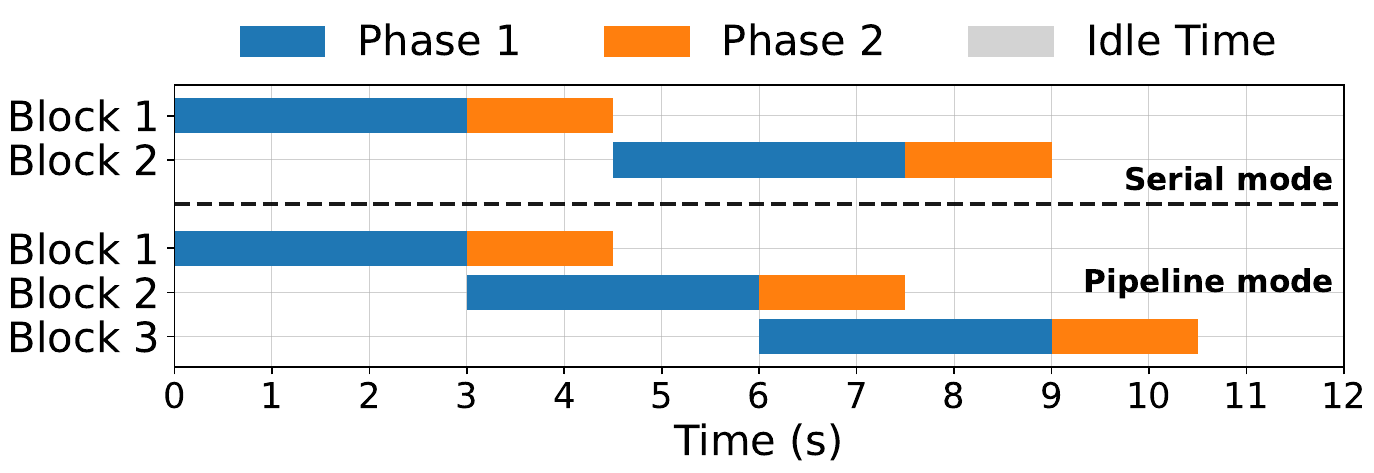}
\label{fig:phase1_dominant}
\end{subfigure}
\hfill
\begin{subfigure}[b]{0.48\textwidth}
\centering
\includegraphics[width=\linewidth]{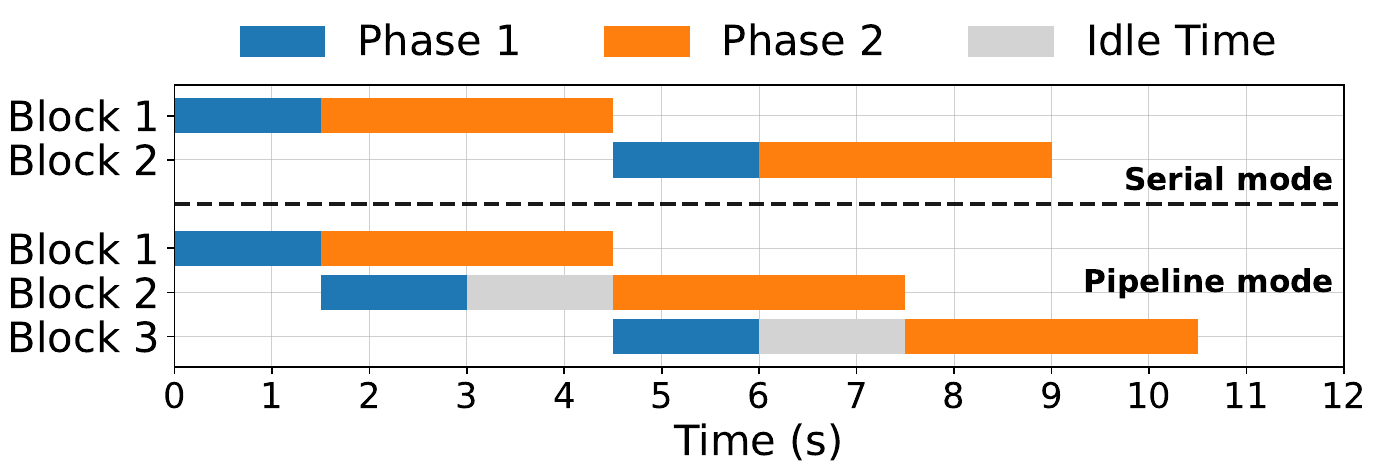}
\label{fig:phase2_dominant}
\end{subfigure}
\caption{Illustration of pipeline bottlenecks caused by imbalanced phase durations. \textbf{Left}: Phase 1 is dominant, causing Phase 2 to be idle. \textbf{Right}: Phase 2 is dominant, causing blocks to wait. Optimal throughput is achieved when the time taken for Phase 1 and Phase 2 are nearly equal so that overlap is maximized.}
\label{fig:pipelining_bottlenecks}
\end{figure*}

\textit{Motivation. }In standard HLF, each peer processes blocks from its commit queue strictly one at a time.
A block must pass through all five commit stages: VSCC validation, private-data fetch, MVCC check, block-store update, and state-database commit, before the peer begins the next block; see Figure~\ref{fig:hlf_validation_phases}.
This strictly serial discipline can leave resources underutilized: while a peer performs sequential ledger writes for one block, available CPU cores cannot simultaneously process validation work for subsequent blocks.
The resulting underutilization becomes the primary commit-phase bottleneck, capping throughput well below what the hardware could sustain.
A natural response is to pipeline consecutive blocks; however, doing so is constrained by dependencies between validation and state-update stages.
The five commit stages are not uniformly parallelizable: VSCC checks and private-data fetches for a block are independent of the preceding block's ledger state, but MVCC validation and ledger writes depend on the state left by the prior block and must therefore execute in strict block order.
Any pipelining scheme must respect this dependency boundary to preserve ledger correctness and determinism.

\textit{Objective. }Our objective is to design a pipelining strategy that maximizes commit throughput and to characterize how the throughput gain varies with the balance between the parallelizable and sequential portions of the commit path.
 
\textit{Approach. }We partition the five commit stages into two phases.
\begin{itemize}[leftmargin=*]
    \item \textbf{Phase~1} comprises VSCC validation and private-data fetch, which are CPU-bound (cryptographic signature verification) and I/O-bound (network or local data retrieval) tasks that depend on the block's contents and endorsement policy, but not on the world state produced by prior blocks.
    \item \textbf{Phase~2} comprises MVCC validation, block-store updates, and state-database commits, involving state-consistency checks and ledger writes that depend on the outcome of the preceding block and must therefore execute sequentially.
\end{itemize}
We define the \emph{time ratio} as the duration of Phase~1 divided by the duration of Phase~2 for a given block, as this ratio determines the effectiveness of pipelining: when the two phases are roughly equal in duration, the overlap between consecutive blocks is maximized and idle time is minimized.
 
Our pipelining mechanism works as follows.
While block~$b_i$ is being processed in Phase~2, the peer concurrently begins Phase~1 for the next block~$b_{i+1}$.
Block~$b_{i+1}$ may enter Phase~2 only after block~$b_i$ has fully completed Phase~2, preserving the sequential state-update invariant required for correctness.
This scheme is safe because chaincode-level endorsement policies are fixed at deployment time and are not influenced by in-flight transactions, eliminating read-after-write hazards 
between Phase~1 of~$b_{i+1}$ and Phase~2 of~$b_i$.
We evaluate this pipelining strategy on both our HLF testbed and in SimPy simulations.
On the testbed, we compare serial and pipelined commit across all four private-data sharing strategies from Section~\ref{methods:pvt-data} to assess how data availability affects phase balance.
In simulation, we additionally vary the number of CPU cores allocated to VSCC (scaling from 24 to 96 simulated cores) to test whether increased VSCC parallelism can further shorten Phase~1 and improve the time ratio.
 
\textit{Expected Impact. }
We expect pipelining to improve commit throughput across configurations by overlapping independent validation work across consecutive blocks and reducing idle pipeline stages.
However, the magnitude of the gain should depend critically on the time ratio.
When Phase~1 is much longer than Phase~2 (as we expect under the reactive $(1,1)$ private-data strategy) the pipeline will be Phase-1-bottlenecked: Phase~2 resources sit idle waiting for the next block to clear Phase~1.
Proactive data-sharing strategies should shrink Phase~1 by removing network fetches, bringing the two phases closer to balance and unlocking larger gains.
Similarly, increasing VSCC parallelism should further shorten Phase~1 and improve balance, but only up to a point.
Once Phase~1 becomes shorter than Phase~2, the bottleneck flips: additional cores yield diminishing returns because Phase~2 now limits the pipeline rate.
We therefore expect peak throughput gains at a \textit{moderate} level of VSCC parallelization, where Phase~1 and Phase~2 durations are approximately balanced, with diminishing returns beyond this point.


\subsection{Strategic Waiting}~\label{methods:strat-wait}

\textit{Motivation. }The Soft Max-Ht leader selection policy introduced in Section~\ref{methods:leader-selecn} improves endorsement parallelism by allowing any peer within~$\tau$ blocks of the maximum height to endorse.
In practice, however, heterogeneous commit speeds across peers can erode this benefit over time.
A peer that commits faster than its neighbors steadily increases its block height, while slower peers accumulate lag.
Once the gap between the fastest and slowest peer exceeds~$\tau$, the slower peers lose endorsement eligibility, and the system collapses into a single-leader regime -- precisely the bottleneck that Soft Max-Ht was designed to avoid.
This collapse is self-reinforcing: the sole leader endorses most transactions and thus retains relevant private data locally, enabling faster commits, while lagging peers incur additional data fetch delays that further widen the height gap.
The result is a widening height gap, reduced endorsement concurrency, smaller blocks entering the orderer (since fewer endorsed transactions flow in per unit time), and degraded throughput.

\textit{Objective. }Our objective is to design a coordination mechanism that prevents this collapse by keeping peers sufficiently synchronized in block height, thereby sustaining a high number of eligible endorsers throughout the run.
We evaluate effectiveness using overall throughput (transactions committed per second) and the fraction of time for which multiple peers remain endorsement-eligible.

\textit{Approach. }We define the following terms.
A \emph{leader} is any peer whose block height equals the current maximum block height across all peers.
A \emph{lagger} is a peer whose block height falls below the current maximum.
The \emph{height gap} between two peers is the absolute difference in their block heights.
We refer to a \emph{single-leader regime} as the state in which exactly one peer remains endorsement-eligible under Soft Max-Ht.
A \emph{boost} refers to a temporary reduction in a lagging peer's effective commit latency, abstracting the reallocation of resources freed by a paused leader (e.g., CPU, disk I/O, or network bandwidth). to assist the lagger (e.g., by prioritizing private-data dissemination to it).

We propose the following mechanism.
At each block-commit event, each peer checks the current height gap between itself and the peer with the lowest block height.
If a leader detects that it is more than $\tau$ blocks ahead of a lagger (but the gap remains below a predefined safety ceiling) it temporarily pauses its own block commits.
Simultaneously, the lagger receives a \emph{boost} (modeled as a reduction in its mean commit time) reflecting the reallocation of the paused leader's resources.
Once the lagger closes the gap to within~$\tau$, normal commit behavior resumes for both peers.

We evaluate this mechanism through a proof-of-concept Python-based simulation
with two peers whose commit times are drawn from exponential
distributions with different means, creating a natural asymmetry.
Endorsement eligibility follows Soft Max-Ht with $\tau = 5$.
Block size is determined dynamically by the number of endorsed transactions present in the orderer queue at each block-cut instant, allowing us to observe how endorsement concurrency feeds back into block formation.
 
\textit{Expected Impact. }We expect that without intervention, the faster peer will pull away within the first few tens of blocks, collapsing the system into a single-leader regime for the remainder of the run.
Under strategic waiting, the periodic pauses should keep both peers within the $\tau$-window for a substantially larger fraction of the run.
This sustained multi-endorser regime is expected to increase the rate of endorsed transactions entering the orderer, produce larger blocks, and improve throughput relative to the vanilla baseline.
However, because the leader sacrifices commit progress during each pause, the gain is bounded: if pauses are too frequent or too long, the leader's own throughput loss may outweigh the benefit of the lagger's catch-up.
We therefore expect the mechanism to be most effective when the commit-speed asymmetry between peers is moderate -- large enough to trigger single-leader collapse without intervention, but small enough that brief pauses suffice for recovery.


\section{Experiments and Results} \label{sec:Exp_And_Results}

In this section, we experimentally evaluate Hyperledger Fabric under the configurations defined in Section~\ref{sec:Methods}, with the goal of validating the phase-level hypotheses introduced in Sections~\ref{methods:pvt-data}--\ref{methods:block-size}. Specifically, we examine how private-data dissemination, block-size selection, and endorsement peer selection affect endorsement behavior, commit performance, and end-to-end throughput and latency.

We begin with controlled micro-benchmarking of private-data dissemination strategies (\S~\ref{exp:pvt-data-dissem}) and block-size selection under varying loads (\S~\ref{exp:blocksize}), grounding their observed behavior in the phase-level expectations outlined earlier.
We then evaluate two proposed optimization mechanisms: two-phase pipelining for block commits (\S~\ref{exp:pipelining}) and strategic waiting (\S~\ref{exp:strategic-waiting}), which explicitly target commit-stage bottlenecks and endorsement parallelism, respectively.

\subsection{Experimental Setup} 
\begin{figure}
    \centering
    \includegraphics[width=0.50\linewidth]{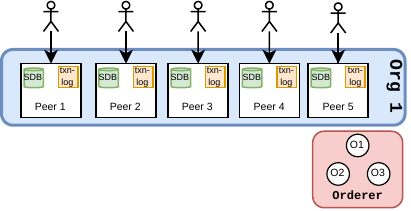}
     \caption{HLF Testbed System Architecture diagram.}
    \label{fig:arch-diagram}
\end{figure}

For both our HLF testbed and SimPy simulations, we use a single-organization HLF deployment (Figure~\ref{fig:arch-diagram}) designed to isolate phase-level performance effects while minimizing cross-organizational confounders.
The setup comprises $5$ peers statically mapped to $5$ clients.

\begin{table}[t]
    \centering
    \caption{Summary of the baseline HLF network configuration.}    
    \label{tab:hlf_common_parameters}
    \begin{tabular}{r|l}
    \hline
        \bf Parameter & \bf Values\\
        \hline\hline
        Number of Organizations & 1\\
         Number of Peers &  5\\
         Number of Clients & 5 \\
         \textcolor{black}{Number of Channels} & \textcolor{black}{1} \\
         Ordering consensus & Raft (3 orderers) \\
         StateDB & \textcolor{black}{LevelDB} \\
        Endorsement concurrency & 10k concurrent requests \\
        Block Size & 4000  \\
        VSCC concurrency & 24 threads \\
        HLF version & HLF v.2.4.7 \\
    \hline
    \end{tabular}
\end{table}

Our HLF testbed experiments are conducted on a production-grade Hyperledger Fabric deployment hosted on a managed Google Kubernetes Engine (GKE) cluster with a Raft-based ordering service. The cluster comprises seven worker nodes in total: three Raft orderer pods co-located on a single node, five client generator pods co-located on a separate node, and five peers placed one per node across the remaining five nodes. Each worker node is provisioned with 24 vCPUs (Intel Broadwell), 32 GB RAM, and 100 GB persistent storage, and runs Google's Container-Optimized OS with \texttt{containerd}.

\begin{table*}[t]
\centering
\caption{Total Time per Endorsement and Commit Phase Across Different Private Data Dissemination Strategies (250 tps for 600 s)}
\label{tab:pvt_data_phase_wise_endorse_commit_times_booktabs}

\resizebox{\textwidth}{!}{%
\begin{tabular}{lccccccccc}
\toprule
\textbf{Setup}
& \multicolumn{2}{c}{\textbf{Endorsement Phase}}
& \multicolumn{6}{c}{\textbf{Validation \& Commit Phase}}
& \textbf{E2E TPS} \\
\cmidrule(lr){2-3}
\cmidrule(lr){4-9}
& \makecell{\textbf{Pvt. Data}\\\textbf{Distr.}}
& \makecell{\textbf{Total}\\\textbf{Endors.}}
& \textbf{VSCC}
& \makecell{\textbf{Pvt. Data}\\\textbf{Fetch}}
& \textbf{MVCC}
& \makecell{\textbf{Block Store}\\\textbf{Pvt Data}}
& \makecell{\textbf{StateDB}\\\textbf{Commit}}
& \makecell{\textbf{Total}\\\textbf{Commit}}
& \\
\midrule
\textbf{1-1}
& $50 \pm 131$
& $227 \pm 344$
& $806 \pm 157$
& $2007 \pm 1216$
& $133 \pm 29$
& $152 \pm 33$
& $824 \pm 125$
& $3925 \pm 1262$
& 1239 \\

\textbf{4-1}
& $57 \pm 84$
& $360 \pm 503$
& $790 \pm 152$
& $468 \pm 107$
& $135 \pm 35$
& $130 \pm 28$
& $1141 \pm 400$
& $2694 \pm 560$
& 1235 \\

\textbf{4-4}
& $44 \pm 44$
& $275 \pm 424$
& $808 \pm 149$
& $515 \pm 114$
& $148 \pm 54$
& $140 \pm 38$
& $1082 \pm 305$
& $2696 \pm 453$
& 1179 \\

\textbf{4-1*}
& $23 \pm 31$
& $182 \pm 317$
& $812 \pm 143$
& $516 \pm 183$
& $141 \pm 57$
& $135 \pm 60$
& $976 \pm 229$
& $2529 \pm 456$
& 1241 \\
\bottomrule
\end{tabular}%
}

\vspace{0.5em}
\footnotesize
All durations are in milliseconds (ms), averaged across 3 separate runs for each setup.
Values are presented as mean $\pm$ standard deviation.

\end{table*}

\subsection{Optimal Private Data Sharing Strategies}\label{exp:pvt-data-dissem}

HLF's private data mechanism poses a performance trade-off choice: when/where to share data. An endorser can share with one peer, i.e., follow the $(1,1)$ rule, yielding fast endorsements, but risks slow commits for other peers through on-demand private data fetches. Conversely, broadcasting ensures the data is locally available for fast commits at all peers, but delays endorsements awaiting multiple slow acknowledgments. 
We quantify these trade-offs using micro-benchmarks on a production-grade HLF deployment (setup in Table~\ref{tab:pvt_data_sharing_hlf_parameters}).

\begin{table}[t]
    \centering
    \caption{Configurations used in Private data sharing experiment}
    \label{tab:pvt_data_sharing_hlf_parameters}
    \begin{tabular}{r|l}
    \hline
       \bf  Parameter & \bf Values\\
        \hline\hline
         Private data sharing & $(1,1)$, $(4,4)$, $(4,1)$ and $(4,1*)$\\
         Transaction Load & 250 txn/sec per peer for 600s\\
        Leader selection & Rank list \\
    \hline
    \end{tabular}
\end{table}

We compare four private data sharing strategies: the \textit{default} $(1,1)$ policy, two \textit{broadcast variants} $(n-1,n-1)$ and $(n-1,1)$, and our proposed $(n-1,1$*$)$ \textit{relaxed quorum} strategy, to answer two questions: (i) How the acknowledgment rule impacts endorsement latency, and (ii) How the dissemination choice creates or removes data fetching bottlenecks during the commit phase. Table~\ref{tab:pvt_data_phase_wise_endorse_commit_times_booktabs} summarizes the results for these four configurations, reporting the average time spent in each stage of the endorsement and the validation/commit phases.\\

\subsubsection{Endorsement Phase Latency}

\begin{figure}[h]
    \centering
    \includegraphics[width=0.5\linewidth]{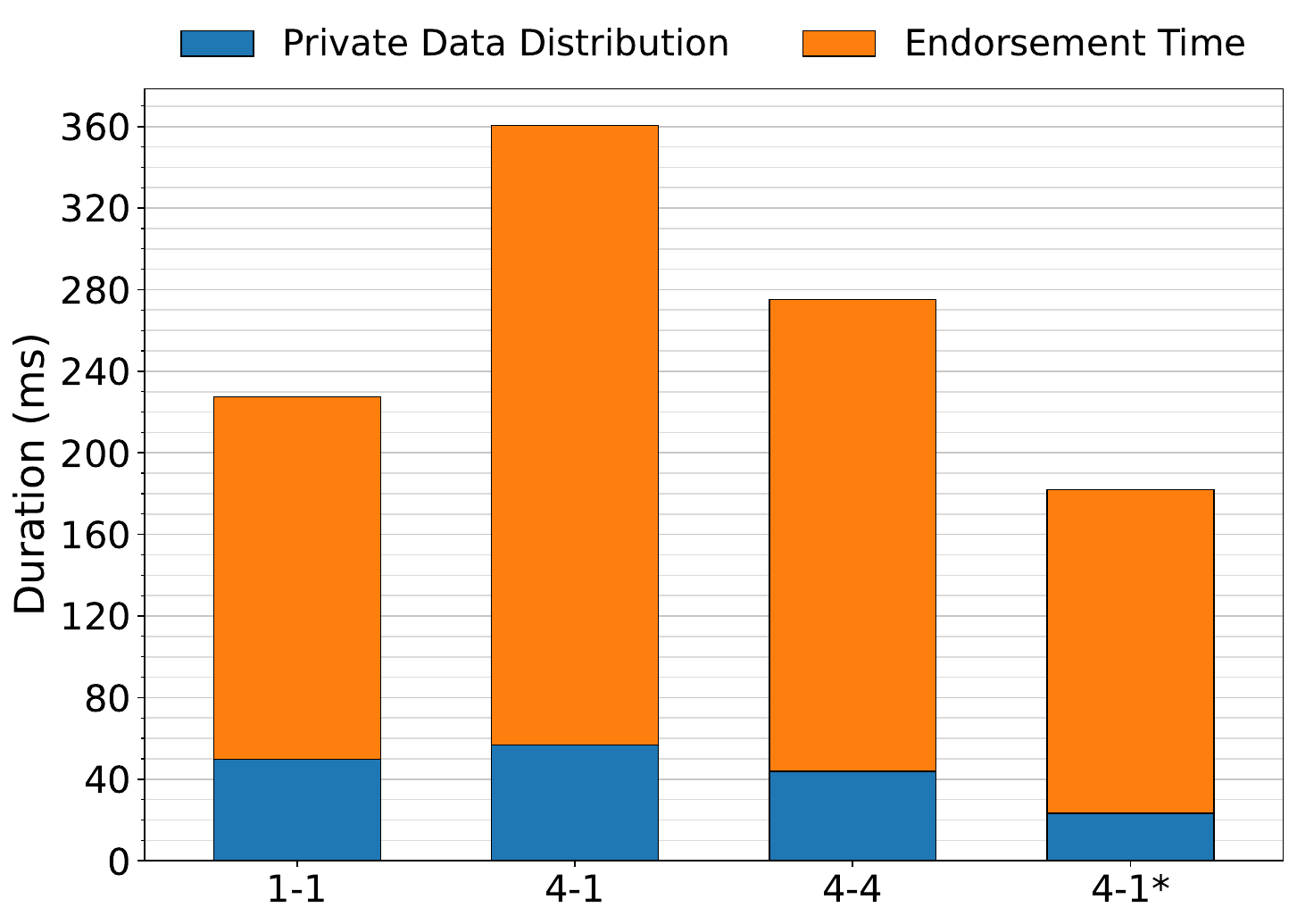}
    \caption{\textbf{Endorsement latency.} Breakdown across private-data dissemination policies. Total latency is dominated by the acknowledgment quorum (single-ack vs all-ack), while private-data distribution time stays similar (3 runs; load $250\times600$).}
    \label{fig:endorse_pvt_data_distribution_avg}
\end{figure}

As shown in Figure~\ref{fig:endorse_pvt_data_distribution_avg}, the time spent distributing private data is surprisingly consistent across all four strategies. Yet, the total endorsement latency varies substantially. This variation is explained by the acknowledgment quorum, i.e., the number of peers an endorser must wait for before proceeding. Under our experimental configuration, strategies that require only a single acknowledgment -- default 
$(1,1)$ and our $(4,1*)$ -- exhibit the lowest endorsement latency. Their average latencies are 227\,ms and 182\,ms (Table~\ref{tab:pvt_data_phase_wise_endorse_commit_times_booktabs} - Total Endors.), thanks to being able to proceed as soon as the first peer responds. In contrast, $(4,4)$ and $(4,1)$ are much slower (275\,ms and 360\,ms) as they are bottlenecked by the slowest responding peer. Also, if all acknowledgments do not arrive before the timeout, they must rerun the full endorsement with different peers, which adds further delays.

\underline{\textbf{Takeaway 1:}} Endorsement latency is dominated by the acknowledgment quorum. Waiting for a single peer is consistently faster than waiting for all acknowledgments.

\begin{figure}[h]
  \centering

  \begin{subfigure}{0.48\linewidth}
    \centering
    \includegraphics[width=\linewidth]{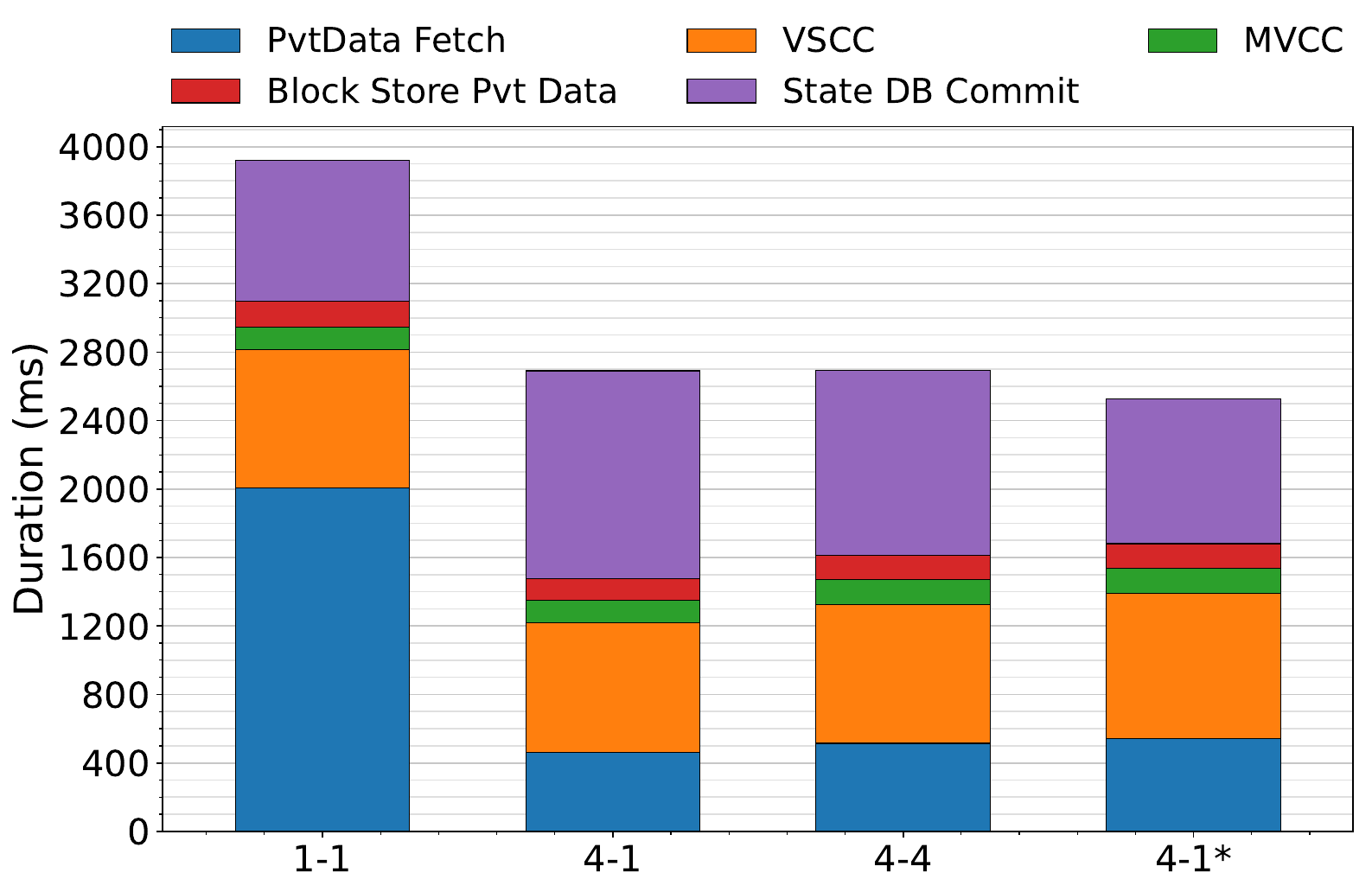}
    \caption{\textbf{Average commit breakdown.} Mean time per commit sub-stage across all peers.}
    \label{fig:PvtData_Avg_mean_commit_bar_plot}
  \end{subfigure}
  \hfill
  \begin{subfigure}{0.48\linewidth}
    \centering
    \includegraphics[width=\linewidth]{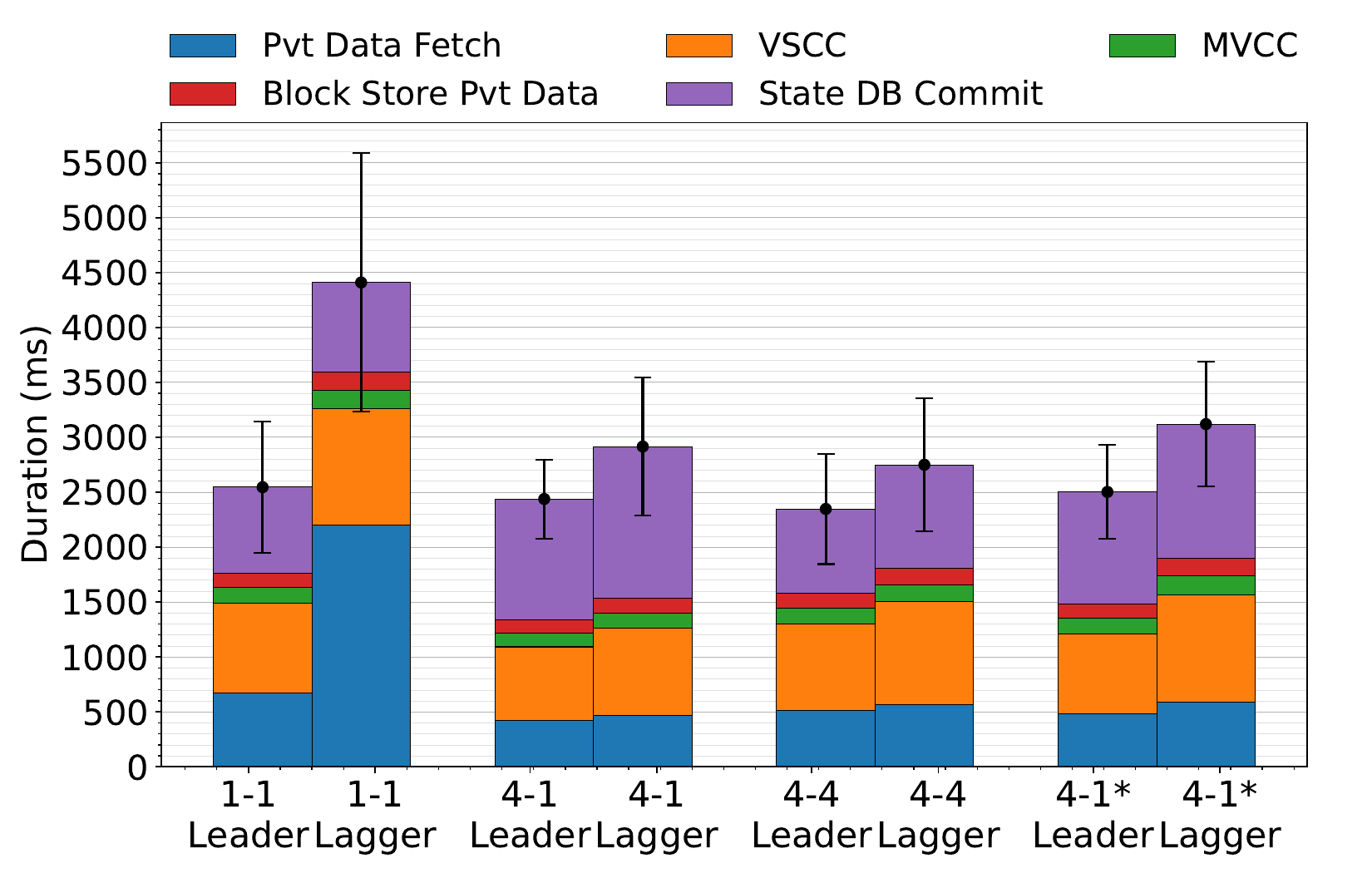}
    \caption{\textbf{Leader vs straggler.} Mean time per commit sub-stage with standard deviation error bars.}
    \label{fig:pvt_data_leader_vs_lagger_bar_plot_with_error_bar}    
  \end{subfigure}

  \caption{\textbf{Commit-phase bottlenecks.} Validation and commit-stage component breakdown for leader vs straggler peers under different private-data dissemination policies.}
  \label{fig:pvtdata-combined}
\end{figure}

\subsubsection{Commit Phase Latency}

While a single-acknowledgment policy excels in the endorsement phase, the performance landscape inverts during the commit phase. Figure~\ref{fig:PvtData_Avg_mean_commit_bar_plot}, which displays the average commit time across all peers, shows that the $(1,1)$ strategy results in the longest commit duration by a significant margin, averaging 3925\,ms (Table \ref{tab:pvt_data_phase_wise_endorse_commit_times_booktabs}). The primary contributor to this high latency is the \textit{Private Data Fetch} stage, which alone accounts for over 2000\,ms on average. In the $(1,1)$ configuration, a transaction's private data is shared with only one other peer besides the original endorser. Consequently, all other peers in the organization are forced to retrieve this data via expensive, time-consuming network calls before they can validate the block, creating a severe bottleneck.

Conversely, the broadcast strategies, $(4,4)$, $(4,1)$ and our proposed $(4,1*)$, exhibit substantially lower commit latency with comparable average durations between 2529-2696\,ms. By proactively disseminating data during the endorsement stage, they ensure it is almost always available in the each peer's local store. This reduces the ``Private Data Fetch'' step to a fast local lookup ($\approx 500$\,ms), effectively removing the network bottleneck that plagues the $(1,1)$ strategy. The minor variations in their commit times can be attributed to differences in local data store retrieval performance.






This performance gap is even more pronounced when differentiating between leader and straggler peers, as shown in Figure \ref{fig:pvt_data_leader_vs_lagger_bar_plot_with_error_bar}. With the $(1,1)$ policy, a vast disparity emerges between the leader's and \textcolor{red}{the} straggler's commit time. Since leader peer(s) endorse most transactions, it always has the private data locally, leading to fast commits. Stragglers, however, must almost always perform network fetches, resulting in significantly higher commit latencies. The broadcast private-data dissemination policies largely mitigate this disparity, maintaining a much tighter and more predictable commit performance across all peers in the network.

\underline{\textbf{Takeaway 2:}} Proactively broadcasting private data during endorsement substantially reduces commit latency by virtually eliminating expensive, on-demand network fetches, particularly for non-endorsing peers.

\begin{tcolorbox}[boxrule=0.8pt, rounded corners, left=2pt, right=2pt, top=6pt, bottom=6pt]
\underline{\textbf{Recommendation}}: Our proposed $(4,1*)$ \textit{relaxed quorum} offers a favorable balance between fast endorsements and predictable commit performance.
\end{tcolorbox}


\subsection{Block Size Selection under Varying Loads}\label{exp:blocksize}

Having examined how private data dissemination affects endorsement and commit latency, we now turn to another global configuration knob: the block size itself. Choosing the block size in an HLF network involves a direct trade-off between transaction latency and throughput, as anticipated in the load-dependent analysis of Section~\ref{methods:block-size}. In this subsection, we quantify this trade-off (setup in Table~\ref{tab:block_size_choice_hlf_parameters}) using two scenarios: a low load (100 TPS) to and a heavy load (800 TPS).

\begin{table}[t]
    \centering
    \caption{Configurations used in HLF Block Size experiment.}
    \label{tab:block_size_choice_hlf_parameters}
    \begin{tabular}{r|l}
    \hline
        Parameter & Values\\
        \hline
         Private data sharing & (4-1)\\
         Transaction Load & 100, 800 txn/sec per peer\\
         Block Sizes & 500, 1000, 1500 and 2000  \\
         VSCC concurrency & 1000 threads \\
        Leader selection & Rank list \\
    \hline
    \end{tabular}
\end{table}

\subsubsection{Low load (100 TPS)} 

Table~\ref{tab:block_size_choice_block_creation_vs_commit_times} and Figure~\ref{fig:block_creation_and_commit_time_stacked_plot} show that the block creation time for different block sizes is significantly larger than the corresponding block commit time. Thus, across all evaluated block sizes, blocks are consistently committed well before the subsequent block is created. Hence, the end-to-end latency of a transaction is influenced mainly by the time it waits in the Orderer queue until the block is created. Consequently, the end-to-end latency of a transaction closely tracks block creation time under this scenario. 
We further observe that, as block size increases, both block creation time and block commit time increase approximately in proportion.

\noindent \underline{\textbf{Takeaway 1:}} Under light load in our experiments, smaller blocks reduce end-to-end latency without affecting throughput.

    

    

\begin{figure}[t] 
    \centering
    
    \begin{subfigure}{0.48\linewidth}
        \centering
        \includegraphics[width=0.95\linewidth]{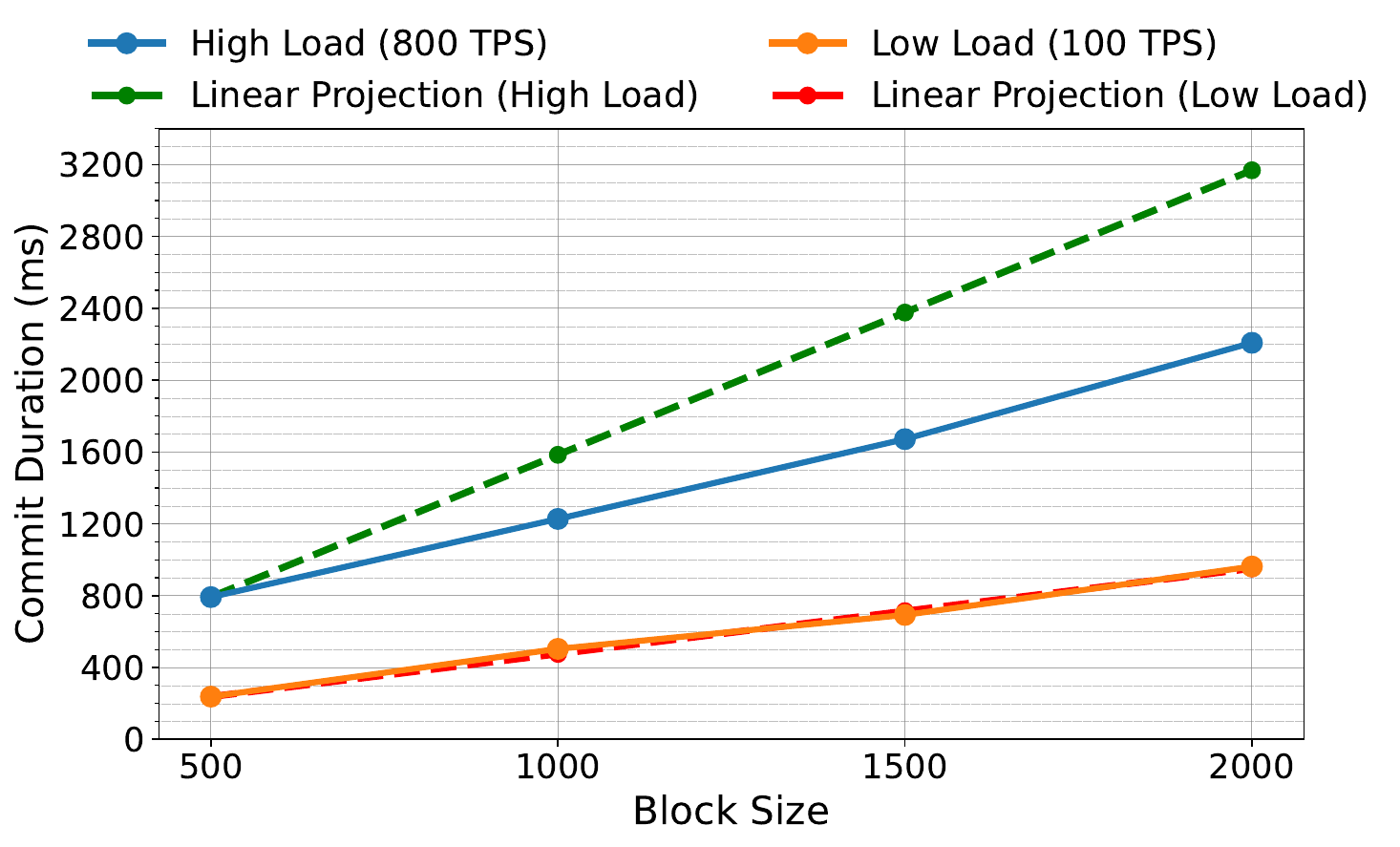}
        \caption{Comparison of average block-commit times under high and low load conditions across different block sizes.}
        \label{fig:block_size_vs_commit_time_line_plot}
    \end{subfigure}
    \hfill
    \begin{subfigure}{0.48\linewidth}
        \centering
        \includegraphics[width=0.95\linewidth]{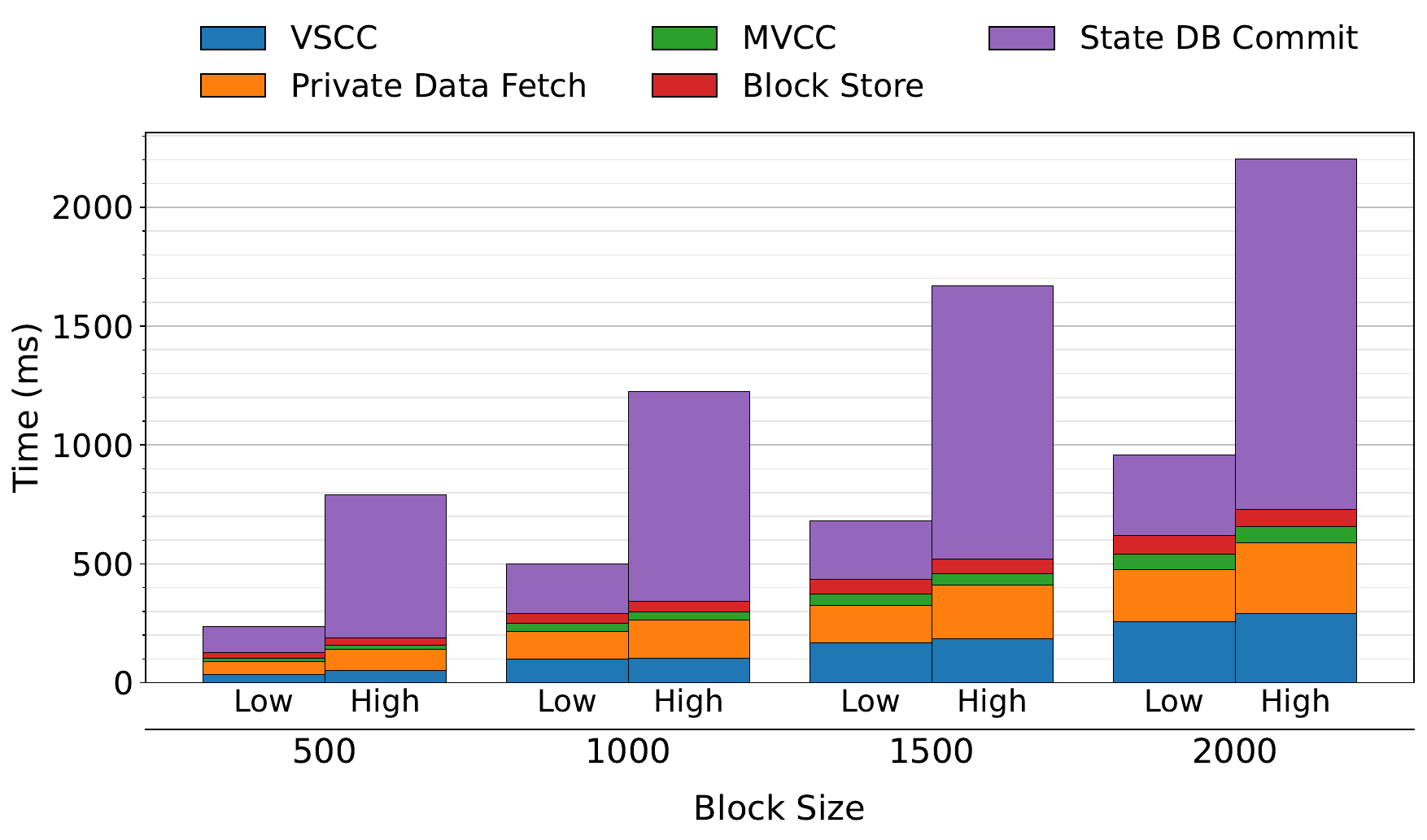}
        \caption{A breakdown of commit phase latency into constituent stages, highlighting that the StateDB Commit stage is the primary bottleneck.}
        \label{fig:commit_phase_breakdown} 
    \end{subfigure}
    
    \caption{Analysis of block commit times under varying loads and block sizes. (a) shows the overall average commit time as load and block size increase. (b) provides a detailed breakdown of the commit phase latency.}
    \label{fig:combined_commit_analysis_vertical} 
\end{figure}

\begin{table*}[h]
    \centering
    \caption{HLF Data. Comparison of Block Size Impact on Performance Metrics under Low and High Load Conditions. Metrics shown include Block Creation Time, Block Commit Time, End-to-End Latency, and Throughput (TPS). E2E is considered for leader only.}  
    \label{tab:block_size_choice_block_creation_vs_commit_times}
    \small

    \resizebox{\textwidth}{!}{%
    \begin{tabular}{@{}c rrrr rrrr@{}}
    \toprule
    \makecell[c]{\textbf{Block} \\ \textbf{Size}}
    & \multicolumn{4}{c}{\textbf{Low Load (100 TPS)}} 
    & \multicolumn{4}{c}{\textbf{High Load (800 TPS)}} \\
    \cmidrule(lr){2-5} \cmidrule(lr){6-9}
    & \makecell[r]{Block Creation \\ Time (s)} 
    & \makecell[r]{Block Commit \\ Time (s)} 
    & \makecell[r]{E2E \\ Latency (s)} 
    & \makecell[r]{E2E TPS} 
    & \makecell[r]{Block Creation \\ Time (s)} 
    & \makecell[r]{Block Commit \\ Time (s)} 
    & \makecell[r]{E2E \\ Latency (s)} 
    & \makecell[r]{E2E TPS} \\
    \midrule
        500  & 1.00 & 0.23 & \textbf{0.98} & \textbf{502.5} & 0.15 & 0.79 & 475.83 & 656.9 \\
        1000 & 2.01 & 0.50 & 1.92 & 503.4 & 0.27 & 1.23 & 349.89 & 847.1 \\
        1500 & 3.00 & 0.68 & 2.22 & 503.8 & 0.40 & 1.67 & 311.85 & 937.5 \\
        2000 & 4.00 & 0.96 & 3.50 & 504.1 & 0.55 & 2.21 & \textbf{305.00} & \textbf{954.9} \\
    \bottomrule
    \end{tabular}%
    }
\end{table*}

\subsubsection{Heavy load (800 TPS)}

\begin{figure}
    \centering
    \includegraphics[width=0.5\linewidth]{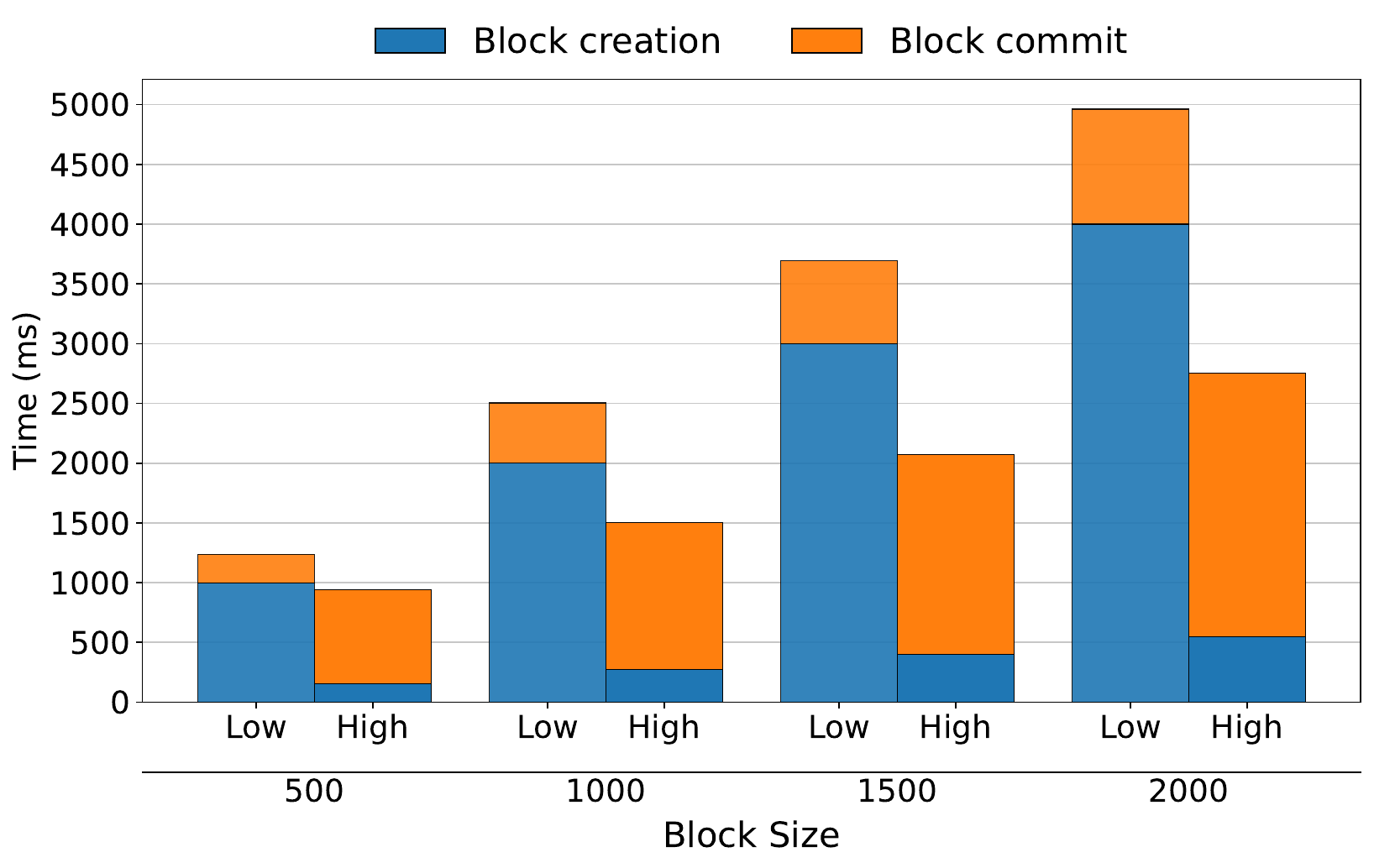}
    \caption{Comparison of average block creation and commit times under high and low load conditions across different block sizes}
    \label{fig:block_creation_and_commit_time_stacked_plot}
\end{figure}

Under heavy load, Table~\ref{tab:block_size_choice_block_creation_vs_commit_times} and Figure~\ref{fig:block_creation_and_commit_time_stacked_plot} show that block commit times are substantially larger than block creation times, which contrasts the low load scenario. Specifically, across all block sizes, a high incoming transaction rate reduces block creation time. On the other hand, as shown in Figure~\ref{fig:commit_phase_breakdown}, commit time increases sharply due to the longer StateDB write time: the latter rises under heavy resource contention from simultaneously endorsing incoming transactions.

This mismatch between block creation and commit rates leads to multiple blocks being created while a prior block is still being committed, causing accumulation in the commit queue.
As a result, end-to-end latency becomes dominated by queuing delay, i.e., the time a block spends in the orderer queue before it is taken up for commit. This explains the large end-to-end latency observed in Table~\ref{tab:block_size_choice_block_creation_vs_commit_times}.

Block commit time, as observed across the evaluated configurations, increases at a slower approximately linear rate as block size grows, as a result of the StateDB write -- the dominant commit component (Figure~\ref{fig:commit_phase_breakdown}) -- showing a similar trend. Figure~\ref{fig:block_size_vs_commit_time_line_plot} illustrates this: committing 2000 transactions as four 500-tx blocks (extrapolated from measured 500-block results) takes about 1.4x longer than committing a single 2000-tx block. This explains the lower end-to-end latency and higher end-to-end throughput for larger blocks.

\noindent\underline{\textbf{Takeaway 2:}} Under heavy load, larger block sizes reduce per-block commit overhead and are observed to improve throughput while lowering overall latency.

\begin{tcolorbox}[boxrule=0.8pt, rounded corners, left=2pt, right=2pt, top=6pt, bottom=6pt]
\underline{\textbf{Recommendation}}: 
Based on our measurements, small block sizes are preferable under low load, while larger block sizes provide better throughput--latency trade-offs under heavy load.
\end{tcolorbox}


\subsection{Criteria for Selecting Endorsement Peers}\label{exp:select-end-peers}

Here, we examine how different endorsement peer selection policies influence HLF performance in terms of endorsement throughput, commit-stage validity and end-to-end transaction success.
 We evaluate this impact using three metrics: the number of transactions (i) successfully endorsed, (ii) dropped due to endorser overload, and (iii) invalidated during the commit phase due to MVCC checks. The last two metrics capture early failures (endorsement drops) and late-stage MVCC invalidations, which incur fundamentally different operational costs and failure semantics.

\begin{table}[h]
    \centering
    \caption{Configurations used in Leader selection experiment}
    \label{tab:leader_selection_parameter_choice}

    \begin{tabular}{l|p{0.42\columnwidth}}
    \hline
        Parameter & Value \\
    \hline
        Private data distribution & (1, 1) \\
        Transaction Load & 250 txn/sec per peer for 300s \\
        gatewayBuffer & 1000 \\
        Endorsement concurrency & 1000 \\
    \hline
    \end{tabular}

\end{table}

By default, HLF employs the Ranked List strategy for endorser selection~\cite{HLF-Docs-Leader-Selection}. For each incoming transaction, peers are ranked in decreasing order of block height, and the first available peer is chosen to endorse. 
In the worst case, this policy may select a peer with a relatively stale ledger state, increasing the likelihood of MVCC invalidations at commit time.
To quantify this effect, we conduct a SimPy-based simulation study. We compare the default Ranked List policy with our proposed Soft Max-Ht strategy (with threshold $\tau = 5$), as well as its two extreme variants: Max-Ht ($\tau = 0$) and All ($\tau = \infty$). Recall that Soft Max-Ht selects only those peers whose block height lies within $\tau$ of the maximum observed block height.

\begin{figure}[h]
    \centering
    \includegraphics[width=0.5\linewidth]{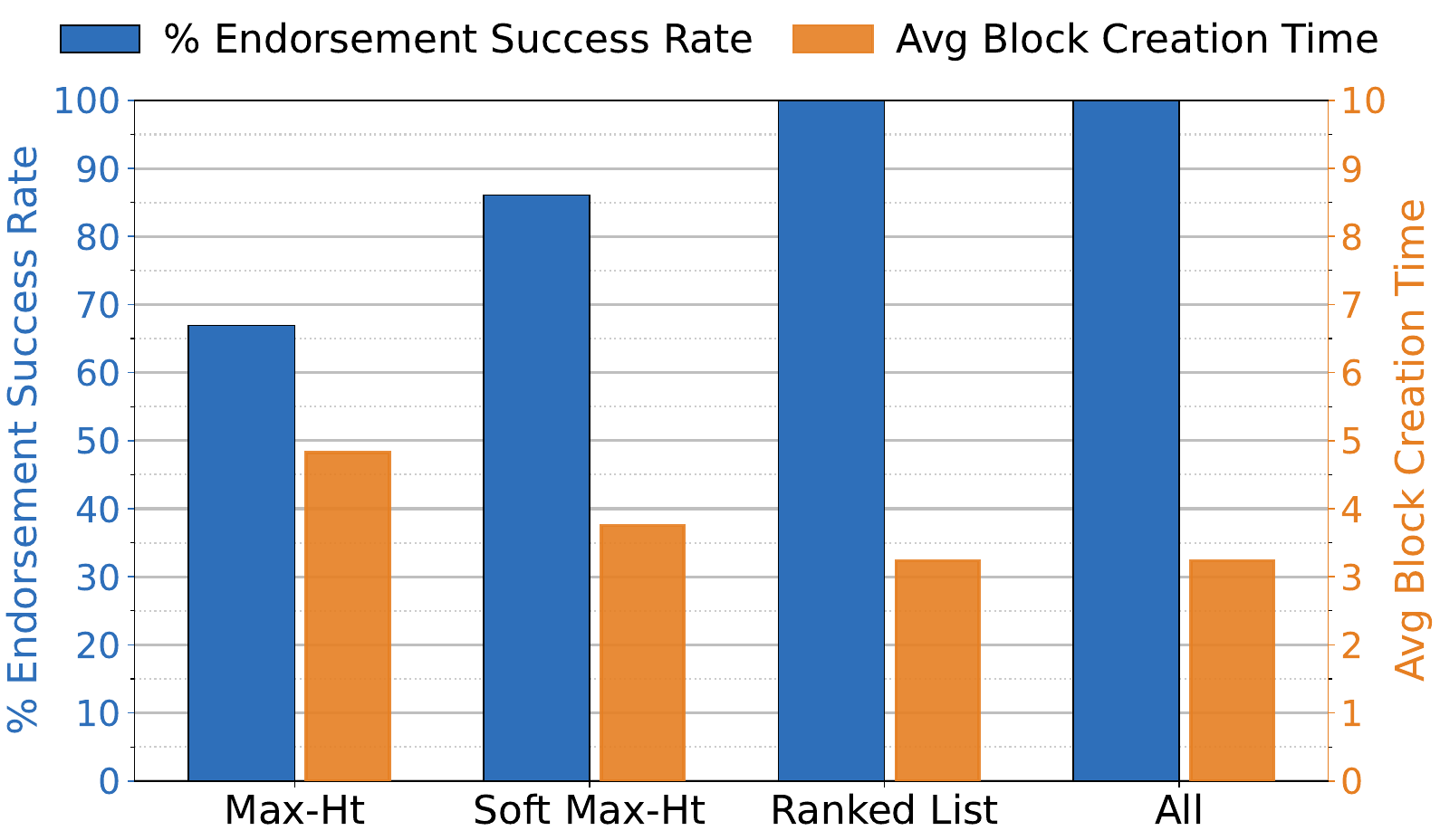}
    \caption{Success rate of transactions and average block creation time for each leader selection strategy.}
    \label{fig:leader_selection_endorsed_transactions}
\end{figure}

The experimental setup is summarized in Table~\ref{tab:leader_selection_parameter_choice}. Each client generates 250 transactions per second over a duration of 300 seconds. The service times for endorsement and commit are sampled from empirical latency distributions obtained from our real-world deployments. At each peer, we configure a gateway buffer of size $1000$ and allow up to $1000$ concurrent endorsements. Thus, an endorser can have at most $2000$ transactions either queued or actively processed at any instant; additional arrivals are dropped. Separately, to model transaction dependencies, an incoming transaction is made dependent on a prior transaction with probability $p$. 
Thus, if a transaction $B$ depends on an earlier transaction $A$ that is still in flight (i.e., not yet committed), $B$ is modeled to fail MVCC validation.
This models read-write and write-write conflicts in which a transaction attempts to access data modified by an in-flight predecessor.
The results are given in Figure~\ref{fig:leader_selection_endorsed_transactions}, Table~\ref{tab:mvcc_inval_ratio_strategies}, and Figures~\ref{fig:leader_selection_success_ratio_heatmap} and~\ref{fig:leader-selection-decision-view-dashboard}. The following observations can be made from them.

\begin{table}[t]
\centering
\caption{SimPy results. Endorsement strategy outcomes across MVCC invalidation ratios. Transaction counts are reported in thousands.}
\label{tab:mvcc_inval_ratio_strategies}
\small
\begin{tabular}{C{1.5cm} l r r r r r}
\toprule

\bf Dep & \bf Strategy & \multicolumn{4}{c}{\bf Number of Transactions} & \bf Success \% \\
\bf Prob & & \em Created & \em Endorsed & \em Dropped & \em Invalid & \\
\midrule
\multirow{4}{*}{0}   & Max-Ht  & $375k$ & 250,957 & 124,043 & 0       & 66.92 \\
                     & Soft Max-Ht     & $375k$ & 322,832 & 52,168  & 0       & 86.09 \\
                     & All          & $375k$ & 375,000 & 0       & 0       & 100.00 \\
                     & Ranked List  & $375k$ & 375,000 & 0       & 0       & 100.00 \\
                     
\midrule
\multirow{4}{*}{0.2} & Max-Ht  & $375k$ & 250,957 & 124,043 & 6,784   & 65.11 \\
                     & Soft Max-Ht     & $375k$ & 322,832 & 52,168  & 18,518  & 81.15 \\
                     & All          & $375k$ & 375,000 & 0       & 33,562  & 91.05 \\
                     & Ranked List  & $375k$ & 375,000 & 0       & 26,887  & 92.83 \\
                     
\midrule
\multirow{4}{*}{0.4} & Max-Ht  & $375k$ & 250,957 & 124,043 & 13,808  & 63.24 \\
                     & Soft Max-Ht     & $375k$ & 322,832 & 52,168  & 36,959  & 76.23 \\
                     & All          & $375k$ & 375,000 & 0       & 67,174  & 82.09 \\
                     & Ranked List  & $375k$ & 375,000 & 0       & 54,018  & 85.60 \\
                     
\midrule
\multirow{4}{*}{0.6} & Max-Ht  & $375k$ & 250,957 & 124,043 & 20,971  & 61.33 \\
                     & Soft Max-Ht     & $375k$ & 322,832 & 52,168  & 55,445  & 71.30 \\
                     & All          & $375k$ & 375,000 & 0       & 100,811 & 73.12 \\
                     & Ranked List  & $375k$ & 375,000 & 0       & 80,920  & 78.42 \\
                     
\midrule
\multirow{4}{*}{0.8} & Max-Ht  & $375k$ & 250,957 & 124,043 & 27,909  & 59.48 \\
                     & Soft Max-Ht     & $375k$ & 322,832 & 52,168  & 74,013  & 66.35 \\
                     & All          & $375k$ & 375,000 & 0       & 134,979 & 64.01 \\
                     & Ranked List  & $375k$ & 375,000 & 0       & 108,309 & 71.12 \\
                     
\midrule
\multirow{4}{*}{1}   & Max-Ht  & $375k$ & 250,957 & 124,043 & 34,651  & 57.68 \\
                     & Soft Max-Ht     & $375k$ & 322,832 & 52,168  & 92,982  & 61.29 \\
                     & All          & $375k$ & 375,000 & 0       & 168,577 & 55.05 \\
                     & Ranked List  & $375k$ & 375,000 & 0       & 135,585 & 63.84 \\
                     
\bottomrule
\end{tabular}
\end{table}

\subsubsection{Impact on Endorsement Success and Block Creation.}

Figure~\ref{fig:leader_selection_endorsed_transactions} shows that stricter policies funnel transactions to fewer peers, leading to saturation and subsequent drops. This effect is most pronounced under Max-Ht, which directs all transactions to the peer with the highest block height. 
With a capacity limit of 2,000 transactions (Table~\ref{tab:leader_selection_parameter_choice}), this peer becomes overloaded under the evaluated workload, resulting in roughly 30\% of proposals being dropped during endorsement.
In contrast, Soft Max-Ht distributes the load among peers close to the maximum height, reducing drops by about 15\%. Under the same workload, more permissive strategies such as Ranked List and All Peers eliminate endorsement drops altogether.

Endorsement drops directly impact block latency. As shown in Figure~\ref{fig:leader_selection_endorsed_transactions}, strategies with fewer drops sustain a steady flow of transactions to the ordering service, resulting in shorter block creation times. In contrast, high drop rates reduce the effective arrival rate at the orderer, slowing block formation and increasing the average block creation time.

\underline{\textbf{Takeaway 1:}} In our simulations, relaxing leader selection reduces endorsement drops and is associated with shorter block creation times.

\subsubsection{Impact on MVCC Invalidations.}

While relaxed strategies improve endorsement success, Table~\ref{tab:mvcc_inval_ratio_strategies} shows that they also lead to higher MVCC invalidations later in the transaction lifecycle. 
This effect stems from stale reads: relaxed policies are more likely to route dependent transactions to peers with outdated world states.

\underline{\textbf{Takeaway 2:}} 
Relaxed leader-selection policies reduce endorsement-stage drops but can increase downstream MVCC invalidations under dependent workloads.

The conditions under which one trade-off is preferable are discussed below.

\subsubsection{Trade-offs: Drops vs. Invalidations}

\begin{figure}[h]
    \centering
    \includegraphics[width=0.5\linewidth]{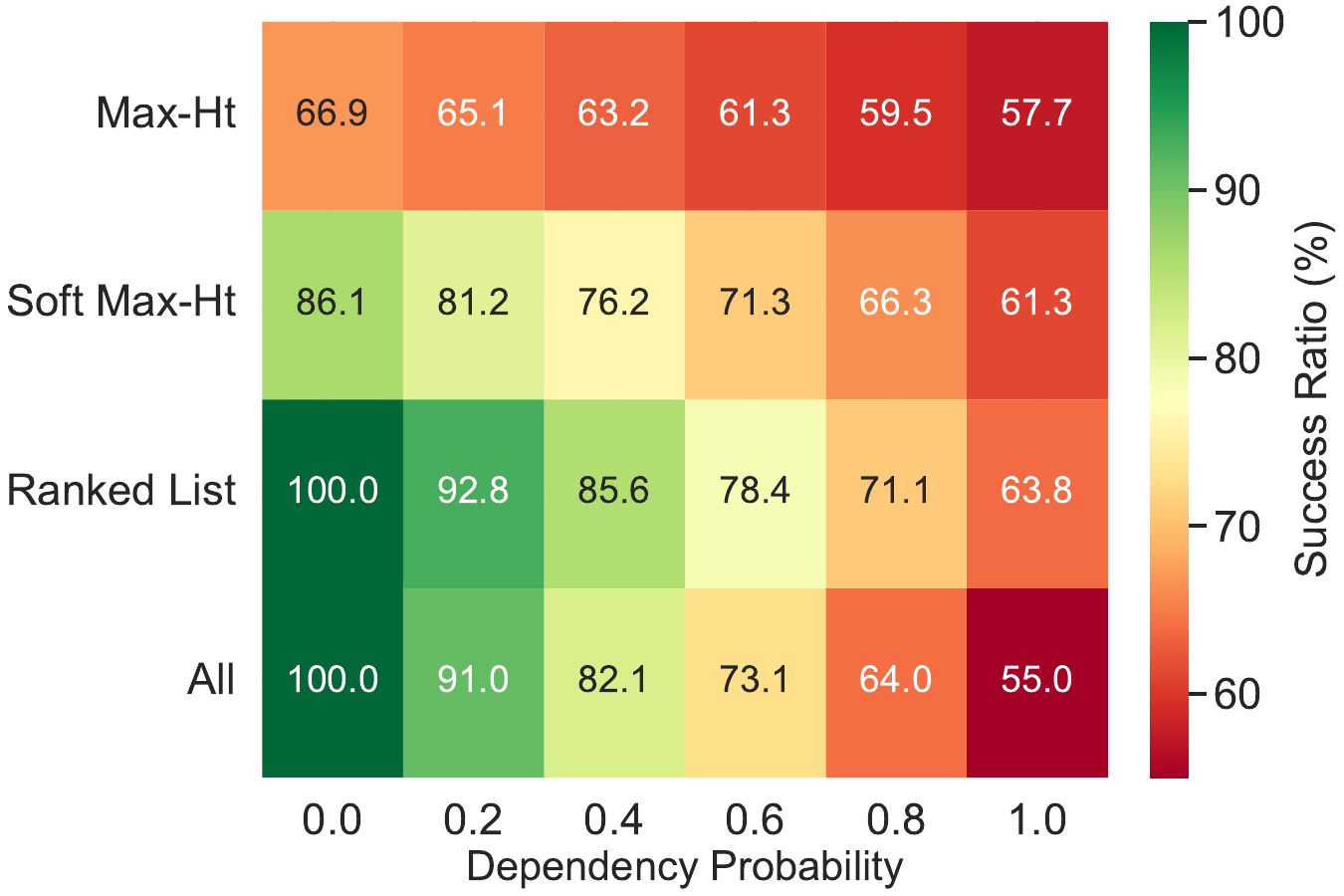}
    \caption{Success Ratio vs Dependency Probability Heatmap for different leader selection strategies.}
    \label{fig:leader_selection_success_ratio_heatmap}
\end{figure}

Figure~\ref{fig:leader_selection_success_ratio_heatmap} presents the transaction success ratio, defined as the ratio of successfully committed \textit{valid} transactions to total submitted transactions as per the numbers reported in Table~\ref{tab:mvcc_inval_ratio_strategies}. Across all transaction-dependency levels ($p$), \emph{Ranked list} achieves the highest success ratio, followed by \emph{All peers}. While both strategies avoid endorsement drops, \emph{Ranked list} outperforms the runner-up because its slightly stricter selection policy leads to fewer MVCC invalidations.

Focusing solely on the success ratio can obscure the operational cost and timing of failures. MVCC invalidations are significantly more expensive than endorsement-stage drops because they consume downstream resources (ordering, replication and validation) and permanently bloat the ledger with invalid entries. In many practical settings, particularly finance-oriented applications~\cite{sharma_fabric_pp}, fail-fast strategies that allow early rejection are preferable to late-stage invalidations. Consequently, a policy with a slightly lower success ratio may still be operationally superior if it reduces costly commit-time failures.

\begin{figure*}[h]
    \centering
    \includegraphics[width=1\linewidth]{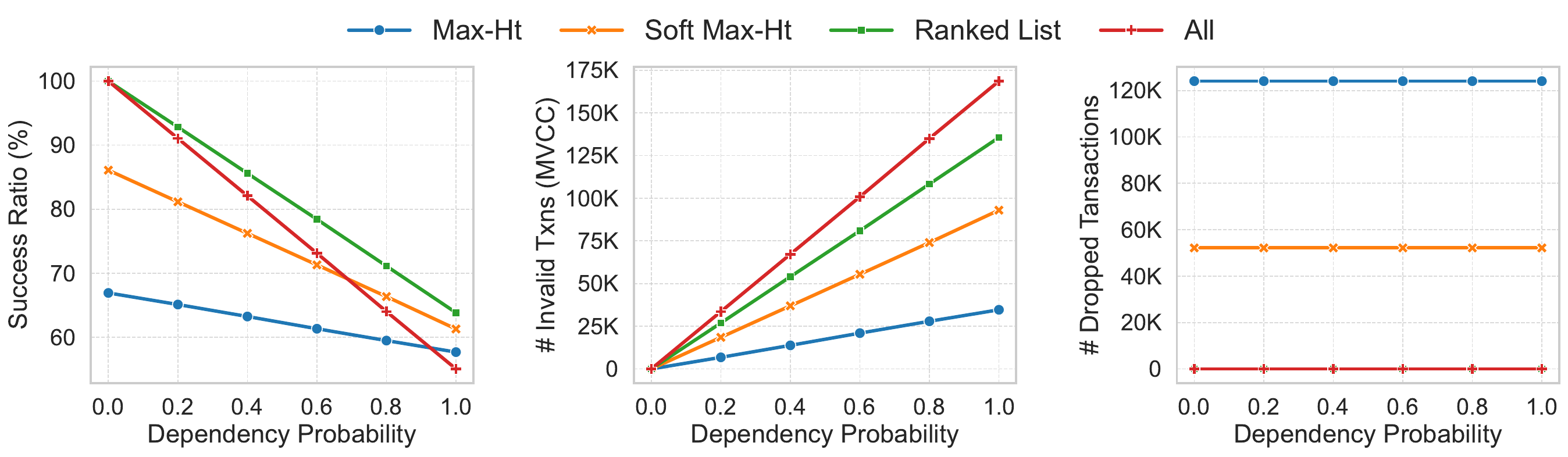}
    \caption{SimPy results showing how increasing MVCC invalidation ratio leads to various trade-offs. (left) success ratio, (middle) MVCC-invalid transactions, and (right) dropped transactions across leader-selection strategies.}
    \label{fig:leader-selection-decision-view-dashboard}
\end{figure*}

Figure~\ref{fig:leader-selection-decision-view-dashboard} reports the success ratio, endorsement-phase drops, and MVCC invalidations for all endorsement policies. For example, at a contention level of $p = 0.4$, multiple strategies achieve a target success ratio of 70\% or higher. Among this feasible set, \emph{Soft Max-Ht} results in fewer MVCC invalidations compared to \emph{Ranked List}. However, this reduction comes at the cost of a higher number of endorsement-phase drops. Therefore, the preferred strategy depends on application requirements. If minimizing MVCC invalidations is critical, \emph{Soft Max-Ht} is more suitable. Conversely, if avoiding endorsement-stage drops is the primary objective, \emph{Ranked List} is preferable.

\begin{tcolorbox}[boxrule=0.8pt, rounded corners, left=2pt, right=2pt, top=6pt, bottom=6pt]
\underline{\textbf{Recommendation:}}
Select a strategy based on the application's failure budget.
If MVCC invalidations must stay low, use \emph{Max-Ht} or \emph{Soft Max-Ht}.
If endorsement drops must stay near zero, use \texttt{Ranked list}.
Avoid \emph{All peers} when contention is non-trivial, since it maximizes MVCC invalidations.
\end{tcolorbox}


\subsection{Two-phase Pipelined Block Commit}\label{exp:pipelining}

Here, we use our HLF testbed and calibrated simulations to empirically evaluate our block-level pipelining strategy (Section~\ref{methods:pipelining}) against the standard serial commit process, comparing end-to-end throughput and latency. Recall that our pipelining strategy overlaps two distinct phases within the commit phase: Phase 1 (VSCC \& private data fetch) and Phase 2 (MVCC \& ledger writes). 

Table~\ref{tab:hlf_common_parameters} summarizes our baseline configuration; unless stated otherwise, all experiments vary a single parameter relative to this baseline.
Unless explicitly noted, end-to-end latency and throughput metrics are reported from the perspective of leader peers, as they represent the critical path for block formation and ordering.

\begin{table}[h]
    \centering
    \caption{Configurations used in our pipelining experiment.}
    \label{tab:hlf_common_parameters_pipelining}
    \begin{tabular}{l|p{0.41\columnwidth}}
    \hline
        \bf Parameter & \bf Values\\
        \hline\hline
        Private data sharing & (1-1), (4-4), (4-1), (4-1*)\\
         Transaction Load &  400 txn/sec per peer for 600s\\
         SimPy Endorser concurrency & 1000 threads \\
         Leader selection & Rank list \\
    \hline
    \end{tabular}
\end{table}

\begin{table*}[h]
    \centering
    \caption{Two-phase pipelined block commit on the HLF testbed (4k block size, 400 txn/s per peer, 600s). We report average block creation time, Phase~1 latency (VSCC + private data fetch), Phase~2 latency (MVCC + ledger writes), and commit throughput under serial vs. pipelined commit across private data sharing setups (1-1, 4-4, 4-1, 4-1*). The performance ratio is the throughput gain of pipelining over serial.}
    \label{tab:block_size_exp_results_400x600}

    \scriptsize
    \setlength{\tabcolsep}{3pt}
    \renewcommand{\arraystretch}{1.1}

    \resizebox{\textwidth}{!}{%
    \begin{tabular}{@{}lccccccccc@{}}
    \toprule
        \multirow{2}{*}{Setup}
        & \multirow{2}{*}{\makecell{Block\\Creation (s)}}
        & \multicolumn{3}{c}{Phase 1 Duration (s)}
        & \multicolumn{1}{c}{Phase 2 Duration (s)}
        & \multirow{2}{*}{\makecell{Time\\Ratio\\}}
        & \multicolumn{2}{c}{Commit TPS}
        & \multirow{2}{*}{\makecell{Performance\\Ratio\\}} \\
        \cmidrule(lr){3-5}
        \cmidrule(lr){6-6}
        \cmidrule(lr){8-9}
        &
        & VSCC
        & \makecell{Pvt. Data\\Fetch}
        & Total
        & \makecell{MVCC + Block Store\\+ StateDB}
        & {$\left(\frac{\text{Phase 1}}{\text{Phase 2}}\right)$}
        & Serial
        & Pipeline
        & {$\left(\frac{\text{Pipeline TPS}}{\text{Serial TPS}}\right)$} \\
    \midrule

        1-1
        & 2.435
        & $2.376 \pm 0.395$
        & $2.028 \pm 0.743$
        & $4.404 \pm 1.138$
        & $1.590 \pm 0.426$
        & 2.769
        & 668.877
        & 908.245
        & 1.357 \\

        4-4
        & 2.365
        & $2.305 \pm 0.395$
        & $0.677 \pm 0.194$
        & $2.983 \pm 0.589$
        & $1.511 \pm 0.373$
        & 1.974
        & 893.351
        & 1340.464
        & 1.500 \\

        4-1
        & 2.581
        & $2.107 \pm 0.509$
        & $0.632 \pm 0.377$
        & $2.740 \pm 0.886$
        & $1.396 \pm 0.455$
        & 1.962
        & 978.887
        & 1447.315
        & 1.478 \\

        4-1*
        & 2.424
        & $2.342 \pm 0.980$
        & $0.828 \pm 2.077$
        & $3.171 \pm 3.058$
        & $1.549 \pm 0.728$
        & 2.056
        & 849.362
        & 1251.065
        & 1.472 \\

    \bottomrule
    \end{tabular}%
    }

    \vspace{0.3em}
    \begin{minipage}{\textwidth}
        \scriptsize
        \emph{Note.} All durations are averages in seconds. Commit TPS is the average transactions per second across all peers.
    \end{minipage}

\end{table*}

\subsubsection{Performance on HLF Testbed}

Table~\ref{tab:block_size_exp_results_400x600} reports Phase~1 and Phase~2 durations and commit throughput (TPS) for serial and pipelined commit across different private data sharing strategies (Section~\ref{methods:pvt-data}). We can make the following two observations from this table.

\begin{enumerate}
    \item \textbf{Pipelining boosts throughput}: Across all private data sharing strategies, pipelining increases commit TPS by 1.36$\times$--1.50$\times$.

    \item \textbf{Commit-TPS improves with proactive data sharing}: The commit TPS is higher for proactive dissemination strategies (4-4, 4-1, 4-1*) due to reduced private data fetch times. This gain is a consequence of Phase 1 and Phase~2 having similar durations, in line with our observations in Figure~\ref{fig:pipelining_bottlenecks}. 
\end{enumerate}

\begin{tcolorbox}[boxrule=0.8pt, rounded corners, left=2pt, right=2pt, top=6pt, bottom=6pt]
\underline{\textbf{Recommendation:}} Use pipelining irrespective of the private data sharing strategy. To maximize gains, use proactive private data sharing strategies, e.g., 4-1, 4-4, or 4-1*.
\end{tcolorbox}

\begin{table*}[h]
    \centering
    \caption{SimPy simulation of CPU core scaling for two-phase pipelined block commit (4k block size). We vary the simulated core count (24--96) by proportionately scaling VSCC latency while sampling private data fetch and Phase~2 (MVCC + ledger writes) latencies from HLF testbed measurements, and report average phase durations, serial vs.\ pipelined commit TPS, and the resulting throughput gain (Pipeline/Serial) across private data sharing setups (1-1, 4-4, 4-1, 4-1*). Values are averages over 3 runs.}
    \label{tab:block_size_exp_results_400x600_vscc_all_cores}

    \scriptsize
    \setlength{\tabcolsep}{3pt}
    \renewcommand{\arraystretch}{1.1}

    \resizebox{\textwidth}{!}{%
    \begin{tabular}{@{}llccccccccc@{}}
    \toprule
        \multirow{2}{*}{Setup}
        & \multirow{2}{*}{\makecell{Sim.\\Cores}}
        & \multicolumn{3}{c}{Phase 1 Duration (s)}
        & \multicolumn{1}{c}{Phase 2 Duration (s)}
        & \multirow{2}{*}{\makecell{Time\\Ratio\\}}
        & \multicolumn{2}{c}{Commit TPS}
        & \multirow{2}{*}{\makecell{Performance\\Ratio\\}} \\
        \cmidrule(lr){3-5}
        \cmidrule(lr){6-6}
        \cmidrule(lr){8-9}
        & & VSCC
          & \makecell{Pvt. Data\\Fetch}
          & Total
          & \makecell{MVCC + Block Store\\+ StateDB}
          & {$\left(\frac{\text{Phase 1}}{\text{Phase 2}}\right)$}
          & Serial
          & Pipeline
          & {$\left(\frac{\text{Pipeline TPS}}{\text{Serial TPS}}\right)$} \\
    \midrule

        \multirow{5}{*}{1-1}
        & 24 & $2.36 \pm 0.37$ & $2.19 \pm 0.16$ & $4.55 \pm 0.53$ & $1.59 \pm 0.44$ & 2.86 & 672  & 891  & 1.32 \\
        & 32 & $1.79 \pm 0.27$ & $2.19 \pm 0.15$ & $3.98 \pm 0.42$ & $1.59 \pm 0.40$ & 2.50 & 739  & 1021 & 1.38 \\
        & 48 & $1.19 \pm 0.21$ & $2.18 \pm 0.14$ & $3.37 \pm 0.35$ & $1.59 \pm 0.43$ & 2.11 & 824  & 1198 & 1.45 \\
        & 64 & $0.88 \pm 0.16$ & $2.19 \pm 0.18$ & $3.07 \pm 0.34$ & $1.59 \pm 0.42$ & 1.93 & 877  & 1307 & 1.49 \\
        & 96 & $0.59 \pm 0.09$ & $2.19 \pm 0.16$ & $2.78 \pm 0.25$ & $1.59 \pm 0.42$ & 1.74 & 937  & 1451 & 1.54 \\

    \midrule

        \multirow{5}{*}{4-4}
        & 24 & $2.29 \pm 0.40$ & $0.67 \pm 0.19$ & $2.97 \pm 0.60$ & $1.51 \pm 0.36$ & 1.96 & 916  & 1359 & 1.48 \\
        & 32 & $1.73 \pm 0.29$ & $0.67 \pm 0.17$ & $2.41 \pm 0.46$ & $1.32 \pm 0.35$ & 1.81 & 1044 & 1671 & 1.60 \\
        & 48 & $1.15 \pm 0.22$ & $0.68 \pm 0.21$ & $1.84 \pm 0.43$ & $1.51 \pm 0.37$ & 1.21 & 1212 & 2186 & 1.80 \\
        & 64 & $0.86 \pm 0.14$ & $0.68 \pm 0.18$ & $1.54 \pm 0.33$ & $1.51 \pm 0.38$ & \textbf{1.01} & 1326 & 2581 & \textbf{1.94} \\
        & 96 & $0.57 \pm 0.11$ & $0.67 \pm 0.19$ & $1.25 \pm 0.30$ & $1.50 \pm 0.34$ & 0.83 & 1468 & 2659 & 1.81 \\

    \midrule

        \multirow{5}{*}{4-1}
        & 24 & $2.12 \pm 0.60$ & $0.64 \pm 0.41$ & $2.76 \pm 1.01$ & $1.40 \pm 0.47$ & 1.96 & 989  & 1459 & 1.47 \\
        & 32 & $1.57 \pm 0.39$ & $0.63 \pm 0.34$ & $2.21 \pm 0.75$ & $1.38 \pm 0.43$ & 1.66 & 1129 & 1822 & 1.61 \\
        & 48 & $1.05 \pm 0.26$ & $0.63 \pm 0.33$ & $1.68 \pm 0.59$ & $1.39 \pm 0.46$ & 1.21 & 1323 & 2387 & 1.80 \\
        & 64 & $0.78 \pm 0.18$ & $0.62 \pm 0.31$ & $1.41 \pm 0.49$ & $1.40 \pm 0.44$ & \textbf{1.00} & 1452 & 2807 & \textbf{1.93} \\
        & 96 & $0.52 \pm 0.12$ & $0.63 \pm 0.38$ & $1.15 \pm 0.51$ & $1.39 \pm 0.47$ & 0.83 & 1589 & 2867 & 1.80 \\

    \midrule

        \multirow{5}{*}{4-1*}
        & 24 & $2.36 \pm 0.83$ & $0.80 \pm 1.57$ & $3.16 \pm 2.41$ & $1.55 \pm 0.71$ & 2.02 & 878  & 1280 & 1.45 \\
        & 32 & $1.79 \pm 0.55$ & $0.76 \pm 1.20$ & $2.53 \pm 1.75$ & $1.56 \pm 0.65$ & 1.61 & 983  & 1588 & 1.61 \\
        & 48 & $1.17 \pm 0.43$ & $0.86 \pm 2.17$ & $2.03 \pm 2.59$ & $1.53 \pm 0.64$ & 1.32 & 1183 & 1972 & 1.66 \\
        & 64 & $0.88 \pm 0.35$ & $0.81 \pm 1.73$ & $1.69 \pm 2.07$ & $1.55 \pm 0.76$ & 1.08 & 1292 & 2332 & 1.80 \\
        & 96 & $0.58 \pm 0.19$ & $0.76 \pm 0.96$ & $1.34 \pm 1.16$ & $1.53 \pm 0.65$ & 0.87 & 1356 & 2480 & 1.82 \\

    \bottomrule
    \end{tabular}%
    }

\end{table*}

\subsubsection{Performance on SimPy Simulations}

Our discussions from the HLF experiments above show that pipelining benefits improve as the durations of the two commit phases become similar. With commit-time private data fetch largely minimized on the testbed, the remaining lever in Phase~1 to maneuver is the VSCC validation step. Since VSCC checks are independent for each transaction and CPU-bound, they are highly parallelizable. This motivates our next hypothesis: allocating more logical CPU resources to VSCC can further improve phase balance and increase commit throughput.

Since our testbed machines had only 24 CPU cores, we use SimPy simulations to test this hypothesis. We first model HLF's end-to-end process in SimPy and simulate hypothetical 32, 48, 64, and 96-core machines. The latency of the processing steps such as Private Data Fetch and Phase 2 durations are obtained by sampling from the empirical data collected from our real-world runs. On the other hand, to get VSCC durations, we proportionately scale the VSCC latency distribution measured on the 24-core testbed under a near-linear speedup assumption, using scaling factors of 0.75 (32 cores), 0.5 (48 cores), 0.375 (64 cores), and 0.25 (96 cores).

\begin{figure}[H]
    \centering

    \begin{subfigure}{0.48\linewidth}
        \centering
        \includegraphics[width=\linewidth]{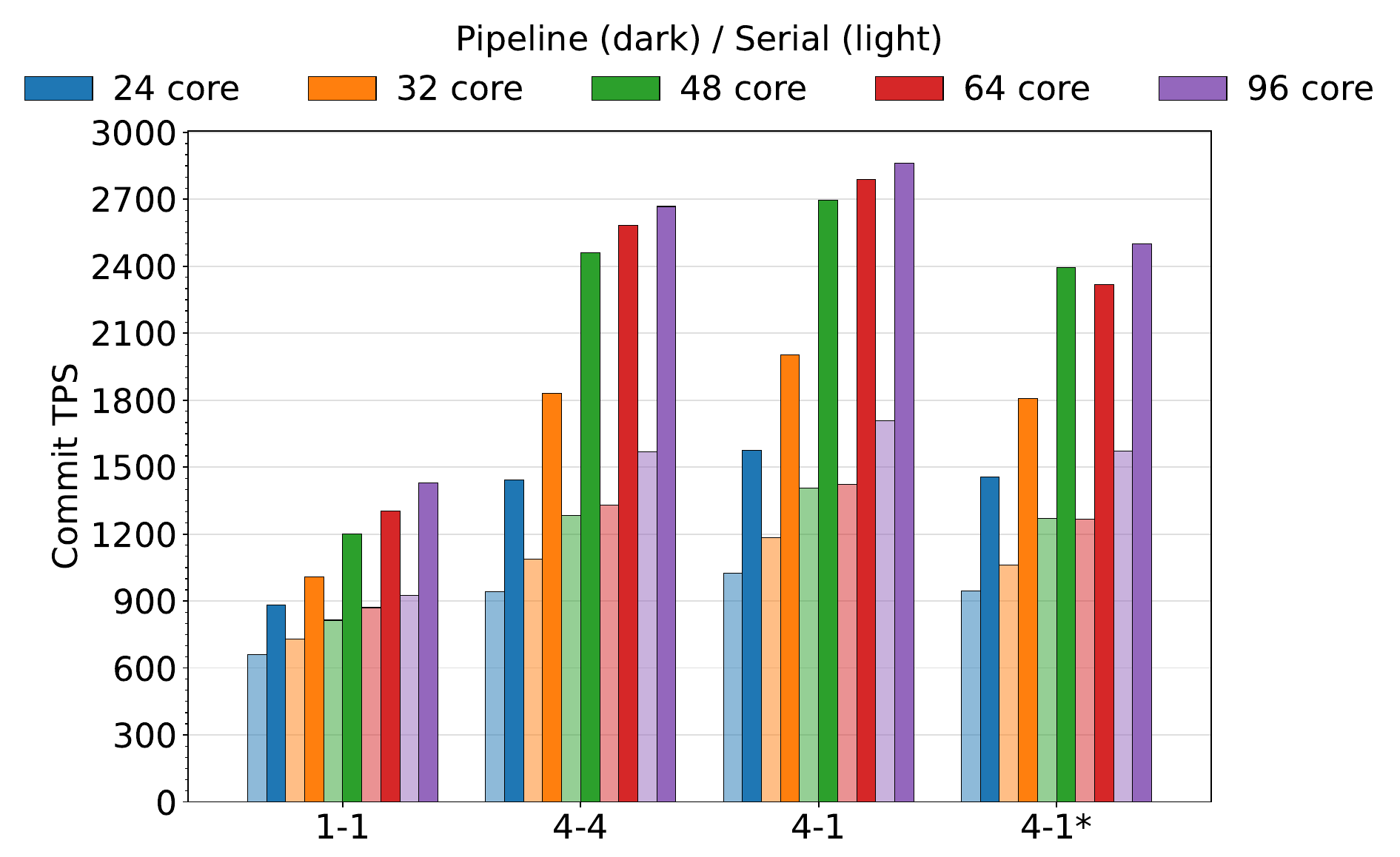}
        \caption{\textbf{SimPy simulation}. Comparison of transaction throughput under simulated VSCC Optimizations across different private data disseminations (1‐1, 4‐4, 4‐1, and 4‐1*) as reported in Table~\ref{tab:block_size_exp_results_400x600_vscc_all_cores}}
        \label{fig:tps_comparison_pipelining-vscc_cores_increase}
    \end{subfigure}
    \hfill
    \begin{subfigure}{0.48\linewidth}
        \centering
        \includegraphics[width=\linewidth]{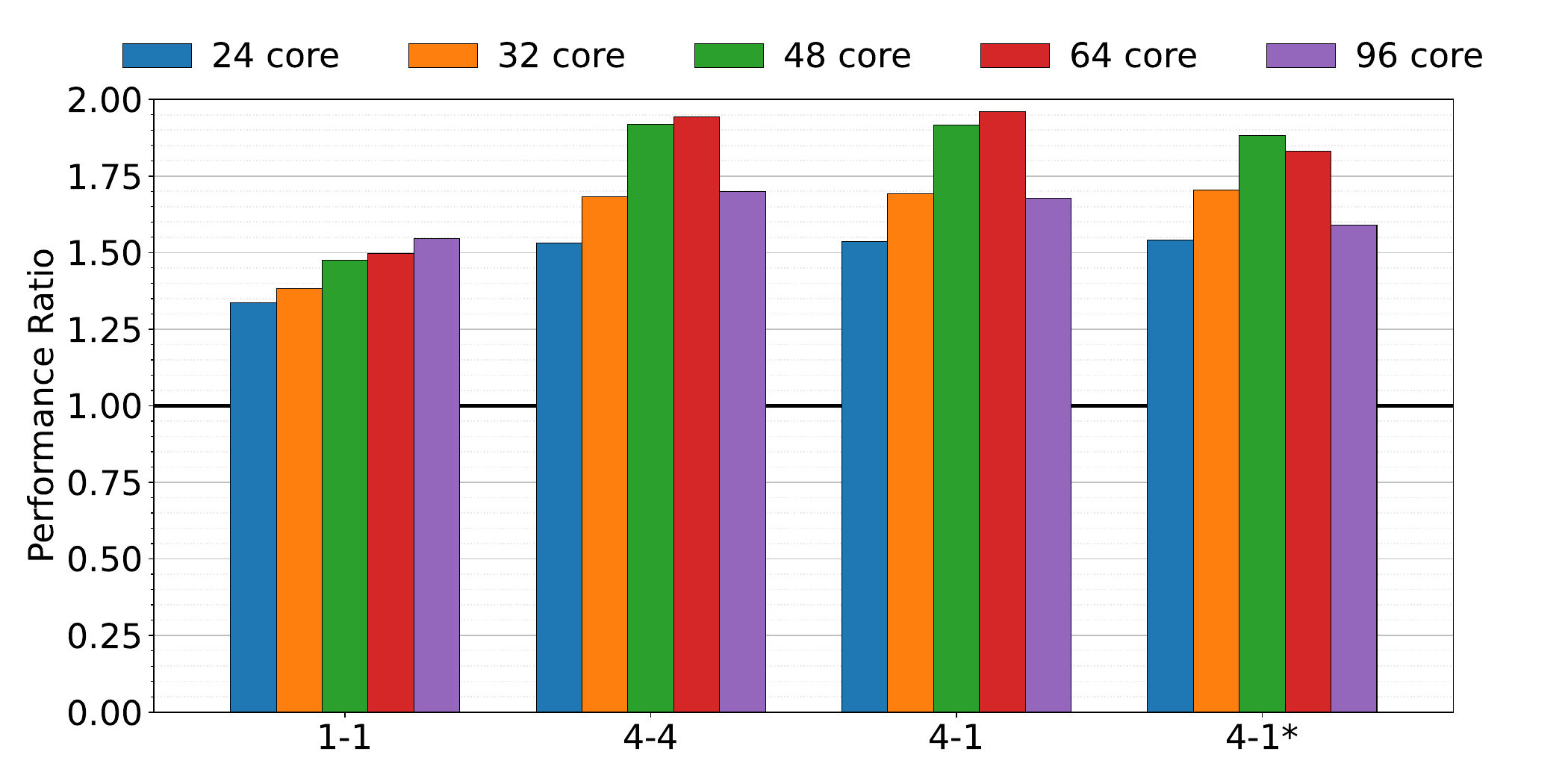}
        \caption{\textbf{Simpy simulation}. Performance ratio of core configurations under Serial and Pipeline modes compared to the actual Serial baseline across different private data disseminations (1-1, 4-4, 4-1, and 4-1*).}
        \label{fig:pipelining_performance_ratio_with_baseline}
    \end{subfigure}

    \caption{\textbf{Impact of VSCC parallelization on pipelined commit performance.} SimPy results comparing transaction throughput and relative performance gains under different simulated VSCC core configurations and private-data dissemination policies.}
    \label{fig:pipelining_vscc_core_scaling_combined}
\end{figure}

The results are summarized in Table~\ref{tab:block_size_exp_results_400x600_vscc_all_cores} and visualized in Figures~\ref{fig:tps_comparison_pipelining-vscc_cores_increase} and~\ref{fig:pipelining_performance_ratio_with_baseline}. Figure~\ref{fig:tps_comparison_pipelining-vscc_cores_increase} shows the resulting commit TPS under serial and pipelined modes, while Figure~\ref{fig:pipelining_performance_ratio_with_baseline} highlights the corresponding performance ratios across core counts. 



Table~\ref{tab:block_size_exp_results_400x600_vscc_all_cores} reveals the following insights: 

\begin{enumerate}
    \item \textbf{VSCC parallelism shortens Phase~1.}: Across all private data sharing strategies, increasing core count reduces VSCC time and thus the total Phase~1 duration.

    \item \textbf{Core scaling initially leads to phase balance and unlocks higher pipelining throughput}: Scaling from 24 to 64 cores shortens Phase~1 toward Phase~2, increasing overlap of the two phases across consecutive blocks and boosting throughput; (4-4) peaks at a 1.94$\times$ gain.

    \item \textbf{Over-scaling flips the bottleneck to Phase~2 and yields diminishing returns}: At 96 cores, Phase~1 becomes much shorter than Phase~2, so Phase~2 limits the pipeline rate and additional cores provide little to no further throughput improvement.
\end{enumerate}

\begin{tcolorbox}[boxrule=0.8pt, rounded corners, left=2pt, right=2pt, top=6pt, bottom=6pt]
\underline{\textbf{Recommendation}}: Combine proactive private data sharing with \textit{just enough} CPU parallelism to make Phase~1 and Phase~2 durations nearly equal. Provisioning beyond this point is inefficient and yields minimal returns.
\end{tcolorbox}


\subsection{Strategic Waiting}\label{exp:strategic-waiting}

As discussed in Section \ref{exp:select-end-peers}, the Soft Max-Ht leader selection policy is preferred in settings where reducing MVCC invalidations is prioritized.
Under this policy, any peer whose block height lies within a configurable threshold $\tau$ of the maximum observed block height is eligible.
So the number of eligible leaders can range between $1$ and $n$, where $n$ is the total number of peers. This enables endorsements to proceed in parallel and potentially improves throughput. 
However, in our simulations, the system often degenerates into a single-leader regime, where only one peer remains endorsement-eligible, eliminating the intended parallelism.

To mitigate this effect, we introduce a strategic waiting mechanism that temporarily suspends commits at the current leader while accelerating lagging peers, allowing them to catch up and re-enter the eligible set.

Table~\ref{tab:strategic_waiting_parameters} summarizes the setup. 
We consider a minimal two-peer setup with a fixed pool of 6000 transactions, chosen to isolate the dynamics of leader--lagger divergence.
Endorsement time is fixed at $0.1$ seconds for both peers. The orderer cuts a block every $2$ seconds, and the resulting block size is determined dynamically by the number of endorsed transactions present in the queue at the cut instant.

\begin{table}[h]
    \centering
    \caption{Configuration used in the strategic waiting experiment.}
    \label{tab:strategic_waiting_parameters}
    \begin{tabular}{l|p{0.30\columnwidth}}
    \hline
        Parameter & Value \\
        \hline
        Number of peers & 2 \\
        Number of transactions & 6000 \\
        Leader selection & Soft Max-Ht ($\tau = 5$) \\
        Commit time distribution means & 1.3s, 2.3s \\
        Endorsement time & 0.1s \\
        Block-cut timeout & 2s \\
        Block size & Dynamic \\
        Boosted lagger mean commit time & 1.8s \\
    \hline
    \end{tabular}
\end{table}

We model Peer 1 and Peer 2 with commit times drawn from $\mathrm{Exp}(\lambda_1)$ and $\mathrm{Exp}(\lambda_2)$, with means given by $1/\lambda_1 = 1.3$s and $1/\lambda_2 = 2.3$s, respectively.
This asymmetry reflects empirical differences observed in private-data handling across peers in HLF: endorsing peers typically have local access to private data and commit faster, whereas non-endorsing peers may incur additional network fetch delays (Figure~\ref{fig:pvt_data_leader_vs_lagger_bar_plot_with_error_bar}). We use the Soft Max-Ht rule for endorsement eligibility, so that any peer whose block height lies within $5$ blocks of the maximum observed block height is treated as an eligible endorser.

\begin{figure*}[h]
    \centering
    \includegraphics[width=\textwidth]{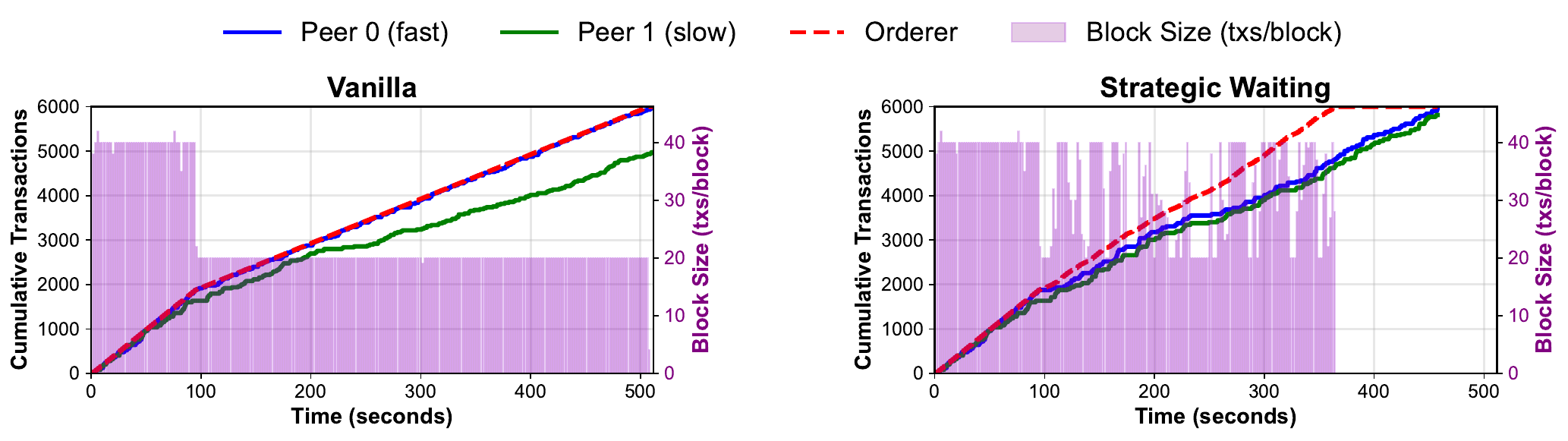}
    \caption{Comparison of cumulative committed transactions for a fast peer (blue), slow peer (green), and the orderer (red) under (Left) Vanilla HLF and (Right) a novel Strategic Waiting strategy. In the Vanilla case, the fast peer leaves the slow peer behind after 100s, leading to smaller, single-endorser blocks and a total duration of over 500s. In the Strategic Waiting case, the leader pauses if more than 5 blocks ahead. This allows the lagger to maintain endorsement eligibility (reflected in variable but frequently larger block sizes) and results in a faster total completion time (<500s) for the same 6,000 transaction workload.}
    \label{fig:sw_heights}
\end{figure*}

We first consider the vanilla case, in which no intervention is applied. Initially, both peers remain close in block height, and hence both are eligible to endorse. Over time, however, Peer 1 commits faster and gradually pulls away from Peer 2. 
Once the gap exceeds the eligibility threshold, Peer~2 is no longer endorsement-eligible, and endorsement load becomes concentrated on Peer~1.
This reduces the rate at which transactions enter the orderer, producing smaller blocks and lower throughput, as seen in Figure~\ref{fig:sw_heights}.

Under strategic waiting, whenever Peer 2 falls behind Peer 1 by more than $5$ blocks and by at most $15$ blocks, 
Peer~1 temporarily suspends block commits, while Peer~2 receives a boost, modeled as a reduction in its mean commit time from $2.3$s to $1.8$s.
This boost can be interpreted as an abstraction of improved assistance from the paused leader, reflecting the leader's ability to reallocate its freed CPU and disk I/O resources toward expediting peer-to-peer data dissemination. Once Peer 2 closes the gap sufficiently, normal commit behavior resumes.

\begin{figure}[h]
    \centering
    \includegraphics[width=0.45\linewidth]{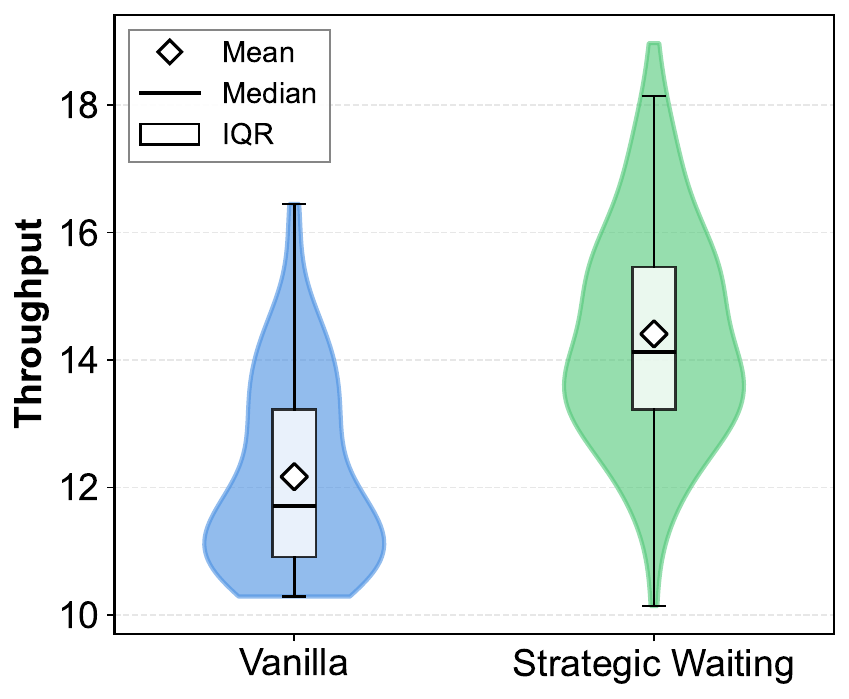}
    \caption{Violin plot comparison of throughput distributions for Vanilla HLF ($\mu$ = \textcolor{black}{12.2} TPS) and Wait-based HLF ($\mu$ = \textcolor{black}{14.4} TPS) systems, showing improved performance with the wait-based approach.}
    \label{fig:sw-throughput-violin}
\end{figure}

By keeping the two peers closer in block height, strategic waiting allows both peers to remain eligible endorsers for a larger fraction of the run. 
This sustains a higher rate of endorsed transactions entering the orderer, which in turn produces larger blocks and faster completion of the 6000-transaction workload in our simulations.
Figure~\ref{fig:sw_heights} illustrates this effect through cumulative committed transactions over time, while Figure~\ref{fig:sw-throughput-violin} shows that the gain persists across repeated runs: the average throughput improves from 12.2~TPS under vanilla HLF to 14.4~TPS with strategic waiting.

\underline{\textbf{Takeaway:}} In our simulations, these initial results suggest that strategic waiting can improve throughput under Soft~Max-Ht by balancing commit progress across peers and preserving endorsement parallelism.

Overall, this mechanism appears promising in simulation, although broader evaluation and integration with production workloads would be required before deployment.

\begin{tcolorbox}[boxrule=0.8pt, rounded corners, left=2pt, right=2pt, top=6pt, bottom=6pt]
\underline{\textbf{Exploratory Recommendation:}}
Strategic waiting may be beneficial in settings where heterogeneous peer commit speeds cause frequent collapse into a single-leader setup. Further evaluation is required before deployment.
\end{tcolorbox}

\section{Conclusions and Future work}\label{sec:Conclusion}

This paper presented a systematic, phase-level study of performance in Hyperledger Fabric (HLF~v2.4.7), focusing on how transaction lifecycle stages interact to shape end-to-end throughput and latency.
Using production-grade experiments and calibrated simulations, we analyzed three key configuration levers -- private-data dissemination, block-size selection, and endorsement peer selection -- and quantified their phase-specific and cross-phase effects.

Our results show that endorsement latency is governed primarily by acknowledgment quorum size, while commit latency is dominated by private-data availability.
Relaxed-quorum private-data dissemination offers a favorable balance between fast endorsements and predictable commit performance.
We further demonstrate that block-size selection is inherently load-dependent: smaller blocks reduce latency under light load, whereas larger blocks amortize commit overhead and improve throughput under heavy load.
For endorsement peer selection, we identify a fundamental trade-off between early endorsement drops and late-stage MVCC invalidations, indicating that policy choice must be guided by application-specific failure tolerance.

Beyond characterization, we proposed two phase-aware optimization mechanisms.
A two-phase pipelined commit overlaps independent validation work across consecutive blocks, improving commit throughput by up to $1.9\times$ when phase durations are balanced.
We also evaluated \emph{strategic waiting}, an exploratory coordination mechanism that mitigates collapse into a single-leader regime under Soft~Max-Ht.
Initial simulation results suggest that it can preserve endorsement parallelism and modestly improve throughput.

There are several important directions to pursue as future work.
Extending these analyses to multi-organization deployments, studying combined effects of multiple optimizations, and designing adaptive controllers for dynamic tuning are promising avenues.
Finally, strategic waiting warrants further evaluation in production deployments to assess its practical viability under realistic workloads and failures.

\bibliographystyle{unsrt}
\bibliography{arXiv-bib}

\end{document}